\documentclass{aa}

\usepackage{color}
\usepackage[dvipsnames]{xcolor}
\usepackage{hyperref}
\hypersetup{colorlinks=true, linkcolor=blue, urlcolor=blue, citecolor=blue}
\usepackage[all]{hypcap}
\usepackage{graphicx}
\usepackage{txfonts}
\usepackage{natbib}
\usepackage{enumitem}
\begin{document} 

   \title{The optimally-sampled galaxy-wide stellar initial mass function}

   \subtitle{Observational tests and the publicly available GalIMF code}

   \author{Z. Yan\inst{1}\fnmsep\thanks{yan@astro.uni-bonn.de}
          \and
          T. Jerabkova \inst{2} \fnmsep \inst{3} \fnmsep \thanks{tereza@sirrah.troja.mff.cuni.cz}
          \and
          P. Kroupa \inst{2} \fnmsep \inst{3} \fnmsep \thanks{pavel@astro.uni-bonn.de}
          }

   \institute{Argelander-Institut f\"{u}r Astronomie, Universit\"{a}t Bonn, Auf dem H\"{u}gel 71, D-53121 Bonn, Germany
         \and
             Helmholtz-Institut f{\"u}r Strahlen- und Kernphysik (HISKP), Universität Bonn, Nussallee 14–16, 53115 Bonn, Germany
         \and
             Charles University in Prague, Faculty of Mathematics and Physics, Astronomical Institute, V Hole{\v s}ovi{\v c}k{\'a}ch 2, CZ-180 00 Praha 8, Czech Republic
             }

   \date{Received April 13, 2017; accepted July 12, 2017}

  \abstract
   {Here we present a full description of the integrated galaxy-wide initial mass function (IGIMF) theory in terms of the optimal sampling and compare it with available observations. Optimal sampling is the method we use to discretize the IMF into stellar masses deterministically. Evidence has been indicating that nature may be closer to deterministic sampling as observations suggest a smaller scatter of various relevant observables than random sampling would give, which may result from a high level of self-regulation during the star formation process. The variation of the IGIMFs under various assumptions are documented. The results of the IGIMF theory are consistent with the empirical relation between the total mass of a star cluster and the mass of its most massive star, and the empirical relation between a galaxy's star formation rate (SFR) and the mass of its most massive cluster. Particularly, we note a natural agreement with the empirical relation between the IMF's power-law index and a galaxy's SFR. The IGIMF also results in a relation between the galaxy's SFR and the mass of its most massive star such that, if there were no binaries, galaxies with SFR $<10^{-4}$ M$_\odot$/yr should host no Type II supernova events.
In addition, a specific list of initial stellar masses can be useful in numerical simulations of stellar systems. For the first time, we show optimally-sampled galaxy-wide IMFs (OSGIMF) which mimics the IGIMF with an additional serrated feature. 
Finally, A Python module, GalIMF, is provided allowing the calculation of the IGIMF and OSGIMF in dependence on the galaxy-wide SFR and metallicity.}

   \keywords{Galaxies: stellar content --
Galaxies: mass function -- Galaxies: star formation -- Star clusters: general -- Galaxies: evolution -- Galaxies: formation}

   \maketitle

\section{Introduction}

The initial distribution of stellar masses as the outcome of star formation is a fundamentally important key for understanding the evolution of stellar systems on star-cluster and galaxy scales. Much effort has been invested in constraining the shape and possible variation of the stellar IMF on the star cluster scale, showing an invariant IMF \citep{Scalo1986,Kroupa2001,Chabrier2003,Bastian2010,Kroupa2013,Hopkins2013}, and on the galaxy scale, suggesting an variant IMF \citep{Salpeter1955,Matteucci1990,Vazdekis2003,Hoversten2008,Meurer2009,Lee2009,Gunawardhana2011}. While the theories describing star formation and IMF are neither fully developed nor understood. A successful theory must be consistent with the observations both on star cluster and on galaxy scales and also it needs to be able to make predictions of new phenomena.

Here we study a particular approach, the IGIMF theory, developed for calculating galaxy-wide IMFs but starting with the IMF constrained observationally on star-cluster scales. With this semi-empirical approach, we can constrain the star-cluster-scale IMF and galaxy-wide IMF and potentially understand the variations of the IMF in different stellar systems.

The IGIMF theory was introduced by \cite{Kroupa2003} where it was called the "field-star IMF". It was originally based on a universal IMF on the star-cluster scale (with later adoption of a systematically varying IMF, see \ref{axiom: IMF} below) and an embedded cluster mass function (ECMF)\footnote{An embedded cluster refers to a correlated star formation event, i.e., a gravitationally-driven collective process of transformation of the interstellar gaseous matter into stars in molecular-cloud overdensities on a spatial scale of about one pc and within about one Myr \citep{Lada2003, Kroupa2013,Megeath2016}. $M_{\mathrm{ecl}}$ is the mass of all stars formed in the embedded cluster.} to generate a variable IGIMF on the galactic scale, leading to a good description in explaining observations, e.g., most recently, \cite{Gargiulo2015, Fontanot2017}. The theory also made the prediction that dwarf galaxies must show a deficit of H$\alpha$ emission relative to UV emission \cite{Pflamm-Altenburg2007,Pflamm-Altenburg2009}. This was verified to be the case \cite{Lee2009}.

It is important to know that there is an additional condition for the IGIMF theory to work. That is, one needs to apply a certain sampling method which determines the stellar masses of a stellar system with a given IMF. It has been shown that the IGIMF theory with random unconditional sampling from an IMF is not consistent with observations (e.g. \citealt{Fumagalli2011,Andrews2013,Dib2017} as further discussed in Appendix~\ref{Appendix_mMrelation}).

Although the shape of the IMF can be estimated by observations to some extent, the sampling procedure that determines the mass of each generated star in a simulation is another issue. Whether it should be random sampling, deterministic sampling, or somewhere in between needs to be clarified. 

Evidence collated in \cite{Kroupa2002}; \cite{Kroupa2013} has been indicating that the simplest interpretation of the IMF, the IMF being a probability distribution function from which stars are randomly sampled, may fail to reproduce the observations as the observed IMF power-law index above a few M$_{\odot}$ shows a scatter smaller than random sampling would give (see \citealt{Kroupa2013} their figure 27). \cite{Hsu2012, Hsu2013} shows in addition that there are star-forming regions with large total stellar mass but a significant deficiency in massive stars, which is inconsistent with a stochastic scenario where 1000 small star clusters of 100 M$_{\odot}$ would statistically generate the same stellar population as a single $10^5$ M$_{\odot}$ star cluster.

In addition, the recent studies from \cite{Stephens2017} and others (see Fig.~\ref{fig:MmaxMecl} below) support a significant empirical correlation of the maximum stellar mass in a cluster and the mass in stars of an embedded cluster (the $m_{\mathrm{str, max}}$--$M_{\mathrm{ecl}}$ relation). The relation was first indicated by \cite{Weidner2006} and was further discussed in \cite{Weidner2013a}. The author argue that the data disagree with the random sampling with a high confidence.

However, there is still a discussion about whether the $m_{\mathrm{str, max}}$--$M_{\mathrm{ecl}}$ relation is consistent with a randomly sampled invariant IMF. Even though it has been shown by \cite{Weidner2014} that the $m_{\mathrm{str, max}}$--$M_{\mathrm{ecl}}$ observations disfavor random sampling, we will discuss this issue in more detail in the future. A further study applying two statistical tests (KS test and standard deviations of the distance between data points) in \citet[in preparation]{Yan2017b} shows that the observations disfavor the scenario of random sampling of stellar masses with a confidence of $4\sigma$.

This paper studies a deterministic method called "optimal sampling" introduced by \cite{Kroupa2013}. With this method the IMF is populated smoothly with no Poisson noise which is favored by observations in comparison with random sampling as will be demonstrated in Sec.~\ref{secsub:m_max-M_ecl} below.

One counter argument against the optimal sampling scenario is that high-mass stars form only in massive parent clusters according to the optimal sampling. But it appears that about 4\% of the O stars are observed in an isolated state. This disagreement has been lifted since \cite{Gvaramadze2012} show that most known isolated O stars are ejected stars related to a parent cluster and the remaining ones that cannot be traced back to their birth cluster, being very rare cases, are possibly due to the two-step ejection mechanism described by \cite{Pflamm-Altenburg2010}. Recently using the HST, \cite{Stephens2017} specifically looked at 7 massive stars thought previously, from Spitzer observations, to be isolated massive stars in the Large Magellanic Cloud (LMC). All of the 7 stars turn out to be within substantial but compact clusters lying on the $m_{\mathrm{str, max}}$--$M_{\mathrm{ecl}}$ relation, which support \cite{Gvaramadze2012}'s conclusion and statistically disfavors the random drawing scenario.

In the present paper, we focus on studying optimal sampling on the galactic scale for the first time. It is worth noting that the mathematical formulation of optimal sampling has been improved by \cite{Schulz2015} (hereafter SPK). We follow the formulation of SPK in this paper even though the original formulation leads to insignificantly different results. The core idea of optimal sampling is not mathematical idealism but to develop a deterministic sampling method that can describe what a self-regulated nature would do to account for the small scatter of the observations. When choosing a sampling method for the numerical initialization of stellar systems, it is most important to consider that the sampled result is consistent with a variety of observations. We will demonstrate later that this is exactly what the OSGIMF model achieves.

The paper is organized as follows: In Section \ref{sec:Model}, the assumptions underlying the IGIMF theory and optimal sampling are described. The python module, GalIMF, we developed to simulate the theory is described in Section \ref{sec: galIMF} and make available for general use\footnote{GalIMF version 1.0.0 is available at https://github.com/Azeret/galIMF. See Sec.\ref{sec: galIMF}}. The results from GalIMF are presented in Section \ref{sec:result} and are compared with observations. Sections \ref{sec:discussion} \& \ref{sec:CONCLUSIONS} contain the discussion and conclusions, respectively.

\section{Model}\label{sec:Model}

\subsection{The IGIMF}\label{sec:Model-IGIMF}

The underlying idea of the IGIMF theory is overly simple and realistic which is that the IMFs assembled in star forming regions must add up to the IMF of the whole galaxy, i.e., first a series of embedded star clusters is generated from the ECMF, then a series of stars for each embedded cluster is generated from the IMF, finally, stars from all the clusters are summed up to generate the galaxy-wide IMF.

To formulate the IGIMF theory, the following five axioms are assumed (see also \citealt{Recchi2015,Fontanot2017}\footnote{We think there are some typos in \cite{Fontanot2017}. In their equation 5, the number -1.06 should be -0.106; In their equation 6, the number 2.35 should be 2.3 and $10^4$ should be $10^6$. In their equation 7, $M_{\mathrm{cl}}$ will be $M_{\rm ecl}$ in our notation.} for comparison) with notation clarified in Table~\ref{table}. Note that we use different symbols for integration variable and object mass. The integration variable applies to the continuous mass distribution function while the object mass accounts to the discretized sampled mass.

\begin{table}
\caption{Notation}
\label{table}
\centering
\begin{tabular}{c c c c}
\hline\hline
Notation & Meaning \\
\hline
    $m$ & integration variable for stellar mass \\
    $m_{\rm max}$ & integration upper limit for stellar mass\\
    $m_{\rm min}$ & integration lower limit for stellar mass\\
    $m_{\rm str}$ & stellar mass \\
    $m_{\rm str,max}$ & most massive stellar mass \\
    $M$ & integration variable for star-cluster mass \\
    $M_{\rm max}$ & integration upper limit for cluster mass\\
    $M_{\rm min}$ & integration lower limit for cluster mass\\
    $M_{\rm ecl}$ & embedded star-cluster mass \\
    $M_{\rm ecl,max}$ & most massive embedded star-cluster mass \\
    $M_{\rm cl}$ & mass of the pre-cluster molecular cloud \\
\hline
\end{tabular}
\end{table}

\begin{enumerate}[leftmargin=0cm,itemindent=0cm,labelwidth=\itemindent,labelsep=0cm,align=left,listparindent=\parindent,label=Axiom \arabic*]
  \setlength\itemsep{1em}
  \item \label{axiom: star form in EC} :
  
  Every star in the galaxy is generated in an embedded cluster.
  
  This assumption is physically plausible because stars form in molecular cloud overdensities which contain much more mass than the mass of the least massive star ($\approx 0.08$ M$_\odot$). This means when a gas cloud collapses to form stars it always forms more than one star, following the particular stellar IMF. Stars in small embedded clusters, e.g., $M_{\mathrm{ecl}}$ less than 50 M$_{\odot}$ or so, will quickly dissolve into the galaxy and may be observed as a distributed young stellar component (e.g. \citealt{Kroupa2003a}). Observations indeed suggest that most and perhaps all the observed stars were formed in embedded clusters \citep{Lada2003, Kroupa2005, Megeath2016}. The properties of galactic-field binary stars also suggest this to be the case \citep{Kroupa1995a, Kroupa1995b}.
  
  \item \label{axiom: IGIMF} :
  
  We define two functions describing (a) the mass distribution of stars that follows the stellar IMF, $\xi_{\mathrm{\star}}(m) = \mathrm{d}N_{\mathrm{\star}}/\mathrm{d}m$, where $\mathrm{d}N_{\mathrm{\star}}$ is the number of stars with mass in the range $m$ to $m+\mathrm{d} m$, and (b) the embedded cluster mass distribution in the galaxy that follows the ECMF, $\xi_{\mathrm{ecl}}(M)=\mathrm{d}N_{\mathrm{ecl}}/\mathrm{d}M$, where $\mathrm{d}N_{\mathrm{ecl}}$ is the number of embedded clusters with mass in stars in the range $M$ to $M+\mathrm{d} M$. Then the stellar mass function for the whole galaxy is the sum of stars in all embedded clusters and can therefore be integrated as:
\begin{equation}\label{eq:xi_IGIMF}
\xi_{\mathrm{IGIMF}}(m,\mbox{SFR})=\int_{0}^{+\infty} \xi_{\mathrm{\star}}(m,M)~\xi_{\mathrm{ecl}}(M,\mbox{SFR})\,\mathrm{d}M.
\end{equation}

This formalism represents a general prescription for the construction of the galaxy-wide IMF from locally valid stellar IMFs. It specifically implies that even if the IMF is universal and fixed on the star-cluster scale, the galaxy-wide IMF can vary. As an extreme example, 1000 small star clusters of 10 M$_{\odot}$ would have the same total mass but a different stellar population than a single $10^4$ M$_{\odot}$ star cluster, simply because all stars more massive than 10 M$_{\odot}$ only exist in the latter case.

  \item  \label{axiom: IMF} :
  
  The stellar IMF, $\xi_{\mathrm{\star}}(m)$, is canonical \citep{Kroupa2001} if the embedded cluster is forming in a gas cloud with a molecular cloud core density $\rho_{\mathrm{cl}}<9.5\times 10^4$ [M$_{\odot}$pc$^{-3}$] (where the molecular cloud core density accounts for the mass of stars and gas as defined in Eq.~\ref{eq:rho_clD} below, following \citealt{Marks2012a}):
\begin{equation}\label{eq:xi_star}
\xi_{\mathrm{\star}}(m,M) =
\begin{cases} 
0, & m<0.08~\mathrm{M}_{\odot}, \\
2k_{\mathrm{\star}} m^{-1.3}, & 0.08~\mathrm{M}_{\odot} \leqslant m<0.5~\mathrm{M}_{\odot}, \\ 
k_{\mathrm{\star}} m^{-2.3}, & 0.5~\mathrm{M}_{\odot} \leqslant m<1\mathrm{M}_{\odot}, \\
k_{\mathrm{\star}} m^{-\alpha_3}, & 1~\mathrm{M}_{\odot} \leqslant m<m_{\mathrm{max}}(M), \\
0, & m_{\mathrm{max}}(M) \leqslant m,
\end{cases}
\end{equation}
where $\alpha_3=2.3$ is the constant Salpeter-Massey index for the invariant canonical IMF but will change for larger $\rho_{\mathrm{cl}}$ to account for IMF variation under star-burst conditions \citep{Elmegreen2003, Shadmehri2004, Dib2007,Dabringhausen2009,Dabringhausen2012, Marks2012a}. 0.08 M$_{\odot}$ in Eq.~\ref{eq:xi_star} is about the lower mass limit of stars \citep{Thies2015}. 

The parameters $k_{\mathrm{\star}}$ and $m_{\mathrm{max}}$ in Eq. \ref{eq:xi_star} are determined simultaneously by solving Eq.~\ref{eq:MeclintMstar} \& \ref{eq:1intMstar} together, i.e., using the mass conservation of the embedded cluster:
\begin{equation}\label{eq:MeclintMstar}
M_{\mathrm{ecl}}=\int_{0.08~\mathrm{M}_{\odot}}^{m_{\mathrm{max}}}m~\xi_{\mathrm{\star}}(m)\,\mathrm{d}m,
\end{equation}
and the optimal sampling normalization condition (which will be explained at the end of \ref{axiom: ECMF}):
\begin{equation}\label{eq:1intMstar}
1=\int_{m_{\mathrm{max}}}^{150~\mathrm{M}_{\odot}}\xi_{\mathrm{\star}}(m)\,\mathrm{d}m,
\end{equation}
where 150 M$_{\odot}$ is the adopted stellar upper mass limit \citep{Weidner2004a, Figer2005, Oey2005, Koen2006, Maiz2007}. Higher mass stars are most likely to be formed through mergers \citep{Banerjee2012a,Banerjee2012b}.

For larger $\rho_{\mathrm{cl}}$, $\xi_{\mathrm{\star}}(m)$ becomes top-heavy where a $\alpha_3(\rho_{\mathrm{cl}})$ relation is adopted from \cite{Marks2012a}:
\begin{equation}\label{eq:alpha_3}
\alpha_3=
\begin{cases} 
2.3, & \rho_{\mathrm{cl}}<9.5\times 10^4, \\
1.86-0.43\log_{10}(\rho_{\mathrm{cl}}/10^6), & \rho_{\mathrm{cl}} \geq 9.5\times 10^4.
\end{cases}
\end{equation}
Here 
\begin{equation}\label{eq:rho_clD}
\rho_{\mathrm{cl}}=3M_{\mathrm{cl}}/4\pi r_{\rm h}^3
\end{equation}
in the unit of $[\mathrm{M}_{\odot}/\mathrm{pc}^3]$ is the pre-cluster molecular cloud core density when the embedded cluster is forming, with $M_{\mathrm{cl}}$ being the original molecular cloud core mass including gas and stars, and $r_{\rm h}$ being the formal half mass radius of the embedded cluster at its theoretical birth time (see \citealt{Marks2012}). Observed binary-star distribution functions constrain this birth density of the cluster.

For a star formation efficiency of 33\%, $M_{\mathrm{cl}}$ is three times the mass of the embedded cluster, $M_{\rm ecl}$. From \cite{Marks2012} we adopt $\log_{10}\rho_{\mathrm{ecl}}=0.61\log_{10}M_{\rm ecl}+2.08$ and $r_{\rm h}/\mbox{pc}=0.1M_{\rm ecl}^{0.13}$, where $M_{\rm ecl}$ is in the unit of [$\mathrm{M}_{\odot}$]. Together with $\rho_{\mathrm{ecl}}=3M_{\rm ecl}/4\pi r_{\rm h}^3$ (Note here the subscript, ecl, is different from Eq.~\ref{eq:rho_clD}, cl.) we have 5 relations and 5 unknown parameters, leading to:
\begin{equation}\label{eq:rho_cl}
\log_{10}\rho_{\mathrm{cl}}=0.61\log_{10}M_{\rm ecl}+2.85.
\end{equation}
Note that the star formation efficiency of 33\% is reasonable but not strict \citep{Banerjee2015,Megeath2016}, and the original molecular cloud core density is not observable. What we realy have here is an empirical relation between $\alpha_3$ and $M_{\rm ecl}$ (combining Eq.~\ref{eq:alpha_3} and \ref{eq:rho_cl}) derived from star cluster observations without a metallicity selection. 

Although Eq.~\ref{eq:alpha_3} only depends on $\rho_{\mathrm{cl}}$, it does not suggest that a fixed Solar metallicity assumption is applied because there is a correlation between the metallicity and $\rho_{\mathrm{cl}}$ (\citealt{Telford2016}\footnote{\cite{Telford2016} conclude a metallicity--SFR relation of galaxies, while SFR correlates with $\rho_{\mathrm{cl}}$ according to our \ref{axiom: ECMF} and \ref{axiom: star formation epoch} below.}). A decomposed empirical relation considering specifically the metallicity environment can also be adopted from \cite{Marks2012a}. We refer to it as the $\alpha_3(x)$ relation:
\begin{equation}\label{eq:alpha_3_metal}
\alpha_3=
\begin{cases} 
2.3, & x<-0.87, \\
-0.41x+1.94, & x>-0.87,
\end{cases}
\end{equation}
where $x=-0.14[\mathrm{Fe}/\mathrm{H}]+0.99\log_{10}(\rho_{\mathrm{cl}}/10^6)$. We note the independent observational evidence for this $\alpha_3(x)$ variation in Eq.~\ref{eq:alpha_3_metal} from a recent analysis of star clusters in M31 \citep{Zonoozi2016, Haghi2017}.

Applying Eq.~\ref{eq:alpha_3_metal} instead of \ref{eq:alpha_3} will affect the high mass end of the IMF and the high mass end of all the following results (Fig. \ref{fig:IGIMF_SFR_} to \ref{fig:alpha3SFR}), but only slightly. Here we choose Eq. \ref{eq:alpha_3} as our fiducial model. Eq. \ref{eq:alpha_3_metal} with fixed Solar metallicity ([Fe/H]=0) being the SolarMetal model will only be considered as a robustness check shown in our final result (Fig. \ref{fig:alpha3SFR}) and Appendix \ref{Appendix}. In general, the fiducial model should be more reliable than a constant metallicity assumption although the difference is not significant for metallicities close to solar value.

With the $\alpha_3$--$\rho_{\mathrm{cl}}$ relation (Eq.~\ref{eq:alpha_3}) and the $\rho_{\mathrm{cl}}$--SFR relation that high SFR galaxies form star clusters reaching to larger masses and thus to higher cloud core densities (see Eq.~\ref{eq:MtotintMecl}, \ref{eq:SFR*deltat} and \ref{eq:rho_cl}), it follows that the IGIMF will become top-heavy for high galaxy-wide SFRs. This is demonstrated in Fig. \ref{fig:IGIMF_SFR_} below.

  \item  \label{axiom: ECMF} :
  
  The ECMF is a single slope power law with a variable power-law index, $\beta$, that depends on the SFR of the galaxy:
\begin{equation}\label{eq:xi_ecl}
\xi_{\mathrm{ecl}}(M,\mbox{SFR})=
\begin{cases}
0, & M<M_{\mathrm{min}},\\
k_{\mathrm{ecl}} M^{-\beta(SFR)}, & M_{\mathrm{min}} \leqslant M<M_{\mathrm{max}}(\mathrm{SFR}),\\
0, & M_{\mathrm{max}}(\mathrm{SFR}) \leqslant M,
\end{cases}
\end{equation}
where $M_{\mathrm{min}}=5$ M$_{\odot}$ is the assumed lower limit of embedded cluster masses roughly corresponding to the smallest stellar groups observed \citep{Kirk2012,Kroupa2003a}, $M_{\mathrm{max}}$ is the upper integration limit in the optimal sampling method defined in SPK and $k_{\mathrm{ecl}}$ is a normalization constant. The determination of these parameters will be detailed below.

Following \cite{Weidner2013b} we assume the ECMF to be flattened (top-heavy) for galaxies with a high SFR:
\begin{equation}\label{eq:beta-SFR}
\beta=-0.106\log_{10}\mbox{SFR}+2,
\end{equation}
where SFR is in the unit of $[\mathrm{M}_{\odot}/\mathrm{yr}]$.

The single slope power law assumption (Eq. \ref{eq:xi_ecl}) is suggested by observation \citep{Lada2003}. The actual shape of the ECMF might be more complicated, e.g. a Schechter-type form that follows the single power law form at the low mass end but turns down rapidly at the high mass end (see \citealt{Lieberz2017}). This kind of modification does not significantly influence the conclusions of the present paper.

The sensitive assumption here is Eq. \ref{eq:beta-SFR}, which directly links the dependency of the IGIMF shape on the SFR (the IGIMF-shape--SFR relation in Fig.~\ref{fig:IGIMF_SFR_} below). The power-law index $\beta$ was observationally found to lie between 1.5 and 2.5 \citep{Weidner2004} which is consistent with Eq. \ref{eq:beta-SFR} for $-5<\log_{10}(\mathrm{SFR}/[\mathrm{M}_{\odot}/\mathrm{yr}])<5$. However, our assumption differs from \cite{Weidner2013b} as they only use Eq. \ref{eq:beta-SFR} for SFR $>1$ M$_{\odot}$yr$^{-1}$ instead of for every SFR. This is because adopting the $\beta$--SFR relation (Eq.~\ref{eq:beta-SFR}) for all SFRs naturally fits the observationally suggested $\alpha_3^{\mathrm{gal}}$--SFR relation for galaxies and our result is actually similar to \cite{Weidner2013b}, which will be shown later in Fig. \ref{fig:alpha3SFR}. The difference in IGIMF shape between applying our and \cite{Weidner2013b}'s $\beta$--SFR relation assumption is shown in Fig.~\ref{fig:IGIMF_SFR_} and Fig.~\ref{fig:alpha3_change_beta_Weidner}, respectively.

Again, the parameters $k_{\mathrm{ecl}}$ and $M_{\mathrm{max}}$ in Eq.~\ref{eq:xi_ecl} are determined by solving Eq.~\ref{eq:MtotintMecl} \& \ref{eq:1intMecl} together, i.e., by invoking the embedded cluster population mass conservation:
\begin{equation}\label{eq:MtotintMecl}
M_{\mathrm{tot}}=\int_{M_{\mathrm{min}}}^{M_{\mathrm{max}}}M~\xi_{\mathrm{ecl}}(M)\,\mathrm{d}M,
\end{equation}
where $M_{\mathrm{tot}}$ is the stellar mass formed within time $\delta t$ that will be introduced in Eq.~\ref{eq:SFR*deltat}, and the optimal sampling normalization condition:
\begin{equation}\label{eq:1intMecl}
1=\int_{M_{\mathrm{max}}}^{10^9~\mathrm{M}_{\odot}}\xi_{\mathrm{ecl}}(M)\,\mathrm{d}M,
\end{equation}
where the upper integration limit of $10^9$ M$_{\odot}$ is the hypothetical physical upper bound of embedded cluster mass roughly corresponding to the limit where star-cluster-type stellar-dynamical systems (ultra compact dwarf galaxies) appear to end (e.g. \citealt{Dabringhausen2008}).

Originally, the normalization form of Eq. \ref{eq:1intMecl} comes from the notion that there is exactly one object in the mass range from $M_{\mathrm{max}}$ to $10^9~\mathrm{M}_{\odot}$. But according to the method of SPK, an embedded cluster is not actually generated in this mass range. So this normalization method is only justified by its ability to describe observations. It can be seen from Fig. \ref{fig:MmaxMecl} \& \ref{fig:SFRMecl} below, respectively, that the generated maximum stellar and embedded cluster mass, $m_{\mathrm{str, max}}$ and $M_{\mathrm{ecl, max}}$\footnote{The meaning of the symbols and notation is listed in Table~\ref{table}.}, fit well with the observational constraints. It is possible to modify the hypothetical upper mass limit or the value on the left-hand side of Eq. \ref{eq:1intMecl} to change the resulting $M_{\mathrm{ecl, max}}$--SFR relation and fit the data but this is not a concern of the present paper.

  \item  \label{axiom: star formation epoch} :
  
  An ensemble of embedded clusters that optimally populate the ECMF is formed in a $\delta t=10$ Myr period with a constant SFR. This is reasonable because, although the short-term fluctuation of the local ($<$ few pc scale) SFR can happen on a dynamical timescale of less than one to a few Myr, a stable long-term galaxy-wide variation typically happens on a timescale of a few hundred Myr (see \citealt{Renaud2016} for interacting galaxies). The period $\delta t$ is called a star formation epoch and we have the following relation:
\begin{equation}\label{eq:SFR*deltat}
M_{\mathrm{tot}}=\mbox{SFR}\cdot \delta t,
\end{equation}
with $M_{\mathrm{tot}}$ being the total stellar mass of embedded clusters formed in $\delta t$ throughout the galaxy.

We note that $\delta t\approx 10$ Myr accounts for the typical observationally deduced galaxy-wide interstellar medium (ISM) time-scale of transforming the ISM via molecular clouds into a new stellar population (\citealt{Egusa2004,Egusa2009}; review: \citealt{Fukui2010}, \citealt{Meidt2015}). The disappearance of large molecular clouds around young star clusters also takes about 10 Myr \citep{Leisawitz1989}. In our model, $\delta t$ was actually determined by the empirical $M_{\mathrm{ecl, max}}$--SFR relation as a larger $\delta t$ will result in a larger $M_{\mathrm{ecl, max}}$, which can violate the $M_{\mathrm{ecl, max}}$--SFR relation. It has been tested in this and previous work (\citealt{Weidner2004}; SPK) that a 10 Myr epoch ensures that the generated embedded clusters reproduce the observed $M_{\mathrm{ecl, max}}$--SFR relation. This consistency between the observationally estimated ISM timescale and the $\delta t$ needed to fit the observational $M_{\mathrm{ecl, max}}$--SFR data is noteworthy and encouraging.

\end{enumerate}

\subsection{The OSGIMF}\label{sec: Model-OSGIMF}

Different from making an integration and resulting in a smooth function with our \ref{axiom: IGIMF}, a sampling procedure will give a discrete list of stellar masses. We use the optimal sampling method from SPK, their equations 1 to 7 \& 9, where their symbol $M_{\mathrm{max}}$ is $M_{\mathrm{max}}$ or $m_{\mathrm{max}}$ here, and $M_{\mathrm{trunc}}$ in their paper is set to be $10^9$ M$_{\odot}$ and 150 M$_{\odot}$ in our Eq. \ref{eq:1intMecl} and \ref{eq:1intMstar}, respectively. We refer readers to SPK for more details of the sampling method. Here we only highlight some important points as SPK made some improvements on the optimal sampling formalism.

In the SPK method, the mass upper integration limit is not the most massive object mass itself but the upper integration limit of the most massive object. This means the mass of the most massive embedded cluster, $M_{\mathrm{ecl,max}}$\footnote{$M_{\mathrm{ecl}}$ stands for the actual embedded cluster mass in star obtained from observation or sampled by our code. It explicitly indicates that $M_{\mathrm{ecl,max}}$ is the most massive object mass instead of an upper integration limit like $M_{\mathrm{max}}$. See Table~\ref{table}.}, in the empirical $M_{\mathrm{ecl,max}}$--SFR relation will not be the upper mass limit $M_{\mathrm{max}}$ in Eq.~(\ref{eq:MtotintMecl}). Rather, $M_{\mathrm{max}}$ is calculated by equation (9) in SPK. The empirical $M_{\mathrm{ecl,max}}$--SFR relation is automatically fulfilled if $\delta t=10$ Myr as stated in \ref{axiom: IMF}. This is shown in Fig.~\ref{fig:SFRMecl} below.

The case of sampling stars within an embedded cluster is similar. The mass of the most massive star in an embedded cluster, $m_{\mathrm{str,max}}$\footnote{$m_{\rm str}$ stands for the actual stellar mass similar to $M_{\mathrm{ecl}}$.}, in the empirical $m_{\mathrm{str,max}}$--$M_{\mathrm{ecl}}$ relation is not used as the upper integration limit, $m_{\mathrm{max}}$, in Eq.~\ref{eq:MeclintMstar}. The SPK optimal sampling formalism with a stellar truncation mass of 150 M$_{\odot}$ automatically fulfills the $m_{\mathrm{str,max}}$--$M_{\mathrm{ecl}}$ relation as is shown in Fig.~\ref{fig:MmaxMecl} below.

The new SPK sampling formalism is more reasonable in the sense that physical mass is only the integration result rather than being the integration upper limit. In addition, the SPK formalism reduces the number of assumptions for the model. The $M_{\mathrm{ecl,max}}$--SFR relation becomes a natural result of optimal sampling instead of an assumption, e.g., \citet[their axiom (5)]{Weidner2013b}.

\section{The publicly available module: GalIMF}\label{sec: galIMF}

A freely available Python code, GalIMF, is developed to compute the IGIMF and the OSGIMF for a given galaxy-wide SFR and metallicity.

The current version, GalIMF version 1.0.0, incorporates the axioms as detailed in Sec.~\ref{sec:Model}. Modifications, e.g., allowing the stellar IMF to vary with metallicity below 1 M$_\odot$ as described in \cite{Marks2012a}, are easy to implement by changing the model number flags.

GalIMF version 1.0.0 developed for the current paper is downloadable at:

https://github.com/Azeret/galIMF

Code usage example, gallery, and future version of GalIMF is available at GalIMF homepage:

https://sites.google.com/view/galimf/home

\section{RESULTS}\label{sec:result}

With GalIMF, we confirm the previous studies that the IGIMF theory, with and without a discretization by sampling procedure, is consistent with the observed $m_{\mathrm{str, max}}$--$M_{\mathrm{ecl}}$, $M_{\mathrm{ecl}}$--SFR and $\alpha_3^{\mathrm{gal}}$--SFR relations, where $\alpha_3^{\mathrm{gal}}$ is the effective power-law index of the galaxy-wide IMF for stars above 1 M$_{\odot}$ and SFR is the galaxy-wide SFR. An inference relation between the galaxy-wide SFR and SNII occurrence is given as a test of the OSGIMF theory.

The OSGIMF, which is the discretized version of the IGIMF, shows additional serrated features that only appear when the deterministic sampling, e.g., optimal sampling\footnote{Optimal sampling as introduced in \cite{Kroupa2013} and improved by \cite{Schulz2015} is not the only possible formalism for a deterministic sampling. There are other deterministic ways to sample a distribution and yield different results, e.g., one could modify the current optimal sampling normalization condition.} with a large ECMF slope of $\beta>2.1$ is applied.

\subsection{$m_{\mathrm{\rm str,max}}$--$M_{\rm ecl}$}\label{secsub:m_max-M_ecl}

\begin{figure}[!hbt]
    \center
    \includegraphics[width=9cm]{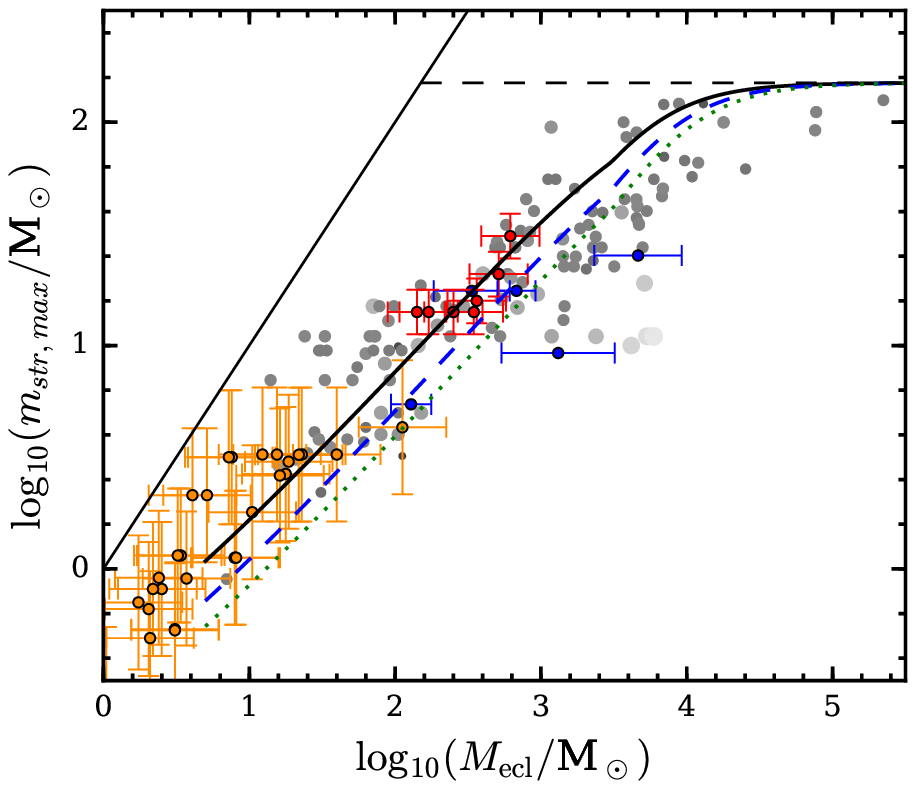}
    \caption{The (most)-massive-stellar-mass--embedded-cluster-mass ($m_{\mathrm{str, max}}$--$M_{\mathrm{ecl}}$) relation. Optimally-sampled result for the most, second and third massive star mass as a function of embedded cluster mass is shown as solid curve, \textcolor{blue}{blue dashed curve} and \textcolor{green}{green dotted curve}, respectively. Observational data come from: \cite{Kirk2012} (\textcolor{orange}{orange dots}), \cite{Stephens2017} (\textcolor{red}{red dots}), \cite{Weidner2013a} (\textcolor{gray}{gray dots}) and \cite{RamirezAlegria2016} (\textcolor{blue}{blue dots}), where \cite{Weidner2013a} is an inhomogeneous set of data culled from the literature for very young clusters without supernova remnants. The average $M_{\mathrm{ecl}}$ uncertainty is 0.34 dex and the average $m_{\mathrm{str,max}}$ uncertainty is 0.16 dex for the gray dots and the gray color is lighter for data points with larger uncertainties. The thin solid line indicates the $M_{\mathrm{ecl}}=m_{\mathrm{str,max}}$ limit and the horizontal thin dashed line indicates the 150 M$_{\odot}$ limit in our Eq. \ref{eq:1intMstar}.}
    \label{fig:MmaxMecl}
\end{figure}

The $m_{\mathrm{str,max}}$--$M_{\rm ecl}$ relation for the optimally sampled result is shown in Fig. \ref{fig:MmaxMecl} to compare with the observations. We plot not only the most massive but also the second and third massive sampled stellar mass. Keep in mind that stellar ejections and mergers \citep{Oh2012, Oh2017} may alter the relation calculated from optimal sampling.

A steepening-feature appears in the $m_{\rm max}$--$M_{\rm ecl}$ relation at $M_{\rm ecl} > 10^{3.49}\,M_\odot$ if the $\alpha_3(\rho_{\rm cl})$ relation (Eq.~\ref{eq:alpha_3}) is used (also shown as thin lines in Fig.~\ref{fig:MmaxMecl23}) when compared to the flatter $m_{\rm str,max}$--$M_{\rm ecl}$ relation at $M_{\rm ecl} > 10^{3.49}\,M_\odot$ if $\alpha_3=$ constant (thick lines in Fig.~\ref{fig:MmaxMecl23}). This steepening, although not so pronounced, is critical to understand a spoon-feature in Fig.~\ref{fig:OSGIMF} that will be discussed in Sec.~\ref{secsub:OSGIMFs} and Appendix~\ref{Appendix2}.

The data points come from different papers. Clusters in \cite{Stephens2017} are estimated to have an age younger than 5 Myr. The average values of "1 Myr estimated" and "2.5 Myr estimated" $M_{\rm ecl}$ in \cite{Stephens2017}'s table 6 are used. The uncertainty of $M_{\rm ecl}$ comes from the estimation of the cluster ages and the resulting $M_{\rm ecl}$ error is approximately from a normal distribution with a $2\sigma$ certainty range of about $\pm 44\%$, or $+0.158/-0.252$ dex according to the description in their section 5.3.. But their $M_{\rm ecl}$ is likely to be underestimated due to extinction and the effect is unclear. Thus we assume an overall error according to a normal distribution in the logarithmic scale with standard deviation $\sigma = 0.126$ dex. The $m_{\rm max}$ values can be overestimated due to the multiplicity of the stars and underestimated due to extinction and the fact that the stars may still be accreting. The $m_{\rm max}$ estimation also suffers several other ambiguous processes and the error was not given (see their section 4.3). Thus we assume an overall error according to a normal distribution in the logarithmic scale with standard deviation $\sigma = 0.13$ dex. From \cite{Kirk2012}, we adopt the top panel of their figure 13. The mass estimations have an uncertainty of order $50\%$ as stated in \citet[their section 2.1]{Kirk2011} which is $+0.176/-0.301$ dex. \cite{Weidner2013a} collate an inhomogeneous set of data culled from the literature for very young clusters without supernova remnants. The catalog contains only clusters with age below 5 Myr to exclude the possibility that the most massive star has already exploded as a supernova and no other selection criterion was applied. The average $M_{\mathrm{ecl}}$ uncertainty is $\pm 0.34$ dex and the average $m_{\mathrm{str,max}}$ uncertainty is $\pm 0.16$ dex for \cite{Weidner2013a}'s data. The uncertainty of the $M_{\rm ecl}$ and $m_{\mathrm{str,max}}$ data points are calculated and the data points are plotted in Fig.~\ref{fig:MmaxMecl} with a different shade of gray, lighter being for larger uncertainties\footnote{The light gray data points are slightly larger for better visibility.}. The blue data points from \citet[their figure 11]{RamirezAlegria2016} are five young clusters but with ages older than the other data presented here. Their ages are roughly up to 10, 7.5, 9, 20 and 7 Myr for data points from left to right. The largest discrepancy for this sample from our model is also the oldest cluster (i.e. 20 Myr) which may already have lost its most massive star though stellar evolution or dynamical ejection. A stellar mass estimation error is not provided by \cite{RamirezAlegria2016}. It is noteworthy how these data follow the dotted line, as expected for such older clusters.

The observational data points in Fig. \ref{fig:MmaxMecl} show no evidence for an intrinsic scatter and support the optimal sampling scenario. This statement is well established in \citet[their figure 1]{Weidner2013a} by showing most of the observational data lie within the region where only 66\% of them should be if stars were randomly sampled from the IMF. A further quantitative discussion disfavoring the random sampling scenario is performed in \citet[in preparation]{Yan2017b} with randomly sampled $m_{\mathrm{str,max}}$--$M_{\rm ecl}$ relations directly compared with the observations. The scatter of the data points is shown to be significantly smaller than random sampling from the IMF and is consistent with no intrinsic scatter given the observational uncertainties.

\subsection{$M_{\mathrm{ecl}}$--SFR}\label{secsub:M_ecl-SFR}

\begin{figure}[!hbt]
    \center
    \includegraphics[width=9cm]{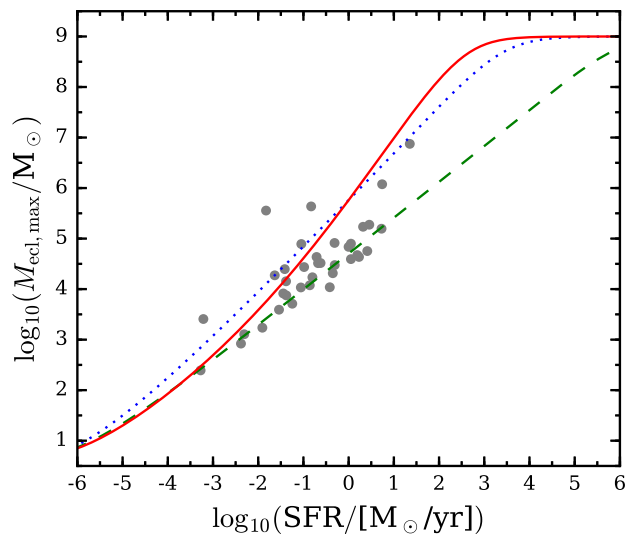}
    \caption{The most-massive young-cluster-mass--galaxy-wide-SFR ($M_{\mathrm{ecl, max}}$--SFR) relation. Optimally-sampled results for different $\beta$ are shown as \textcolor{red}{red solid curve} for $\beta$ following Eq. \ref{eq:beta-SFR}, \textcolor{blue}{blue dotted curve} for $\beta=2$, \textcolor{green}{green dashed curve} for $\beta=2.4$. Observational data (gray dots) adopted from \cite{Weidner2004} have a typical uncertainty of 0.3 dex. The $\beta=2$ and 2.4 curves are almost identical with \cite{Weidner2004}'s figure 6 (middle dotted and middle dashed curves, respectively) that also adopt the $\delta t =10$ Myr assumption. See also \cite{Randriamanakoto2013}.}
    \label{fig:SFRMecl}
\end{figure}

The $M_{\mathrm{ecl}}$--SFR relation for our sampled result is shown in Fig. \ref{fig:SFRMecl}. The data points provided by \cite{Weidner2004} comprise a homogeneous data set of galaxies with young star clusters. The uncertainties of these data points are about 0.5 dex and 0.2 dex for cluster mass and SFR respectively. The cluster masses in this data set are calculated from a magnitude-mass relation assuming the fixed canonical IMF. So this data is consistent with the present analysis when SFR $<0.1$ M$_{\odot}$/yr. For higher SFRs, our $|\alpha_3|$ becomes smaller according to Eq. \ref{eq:alpha_3} (this can be seen explicitly from our Fig. \ref{fig:OSalphaMstar} where the lines for SFR $>10^{-1}$ M$_{\odot}$/yr drop down at $\log_{10}m/\mathrm{M}_{\odot}=0$). The $M_{\mathrm{ecl}}$ values derived with our $\alpha_3$ assumption would be smaller than the data adopted from \cite{Weidner2004} with their canonical $\alpha_3$ assumption. It is possible that the additive constant "2" in Eq.~\ref{eq:beta-SFR} is larger which would fit the data points better, i.e., closer to the green dashed line rather than the blue dotted line (and also makes the "fiducial model" higher in Fig.~\ref{fig:alpha3SFR} below). \cite{Randriamanakoto2013} extend the observation of the empirical relation between the brightest cluster magnitude in a galaxy and the host SFR to higher SFRs. As the $\alpha_3$--SFR dependence makes the $M_{\mathrm{ecl}}$ determination from magnitude very complicated, this data set is not added into our plot. But \cite{Randriamanakoto2013} explicitly note that the small scatter of their data is inconsistent with random sampling from the ECMF, corroborating on the conclusion reached using different data by \cite{Pflamm-Altenburg2013}.

The data points in Fig. \ref{fig:SFRMecl} again display a small scatter which is consistent with the scenario of a highly self-regulated behavior on galaxy scales (see \citealt{Disney2008} and \citealt{Kroupa2015} for a relevant discussion). In addition, our result suggests that a variable $\beta$ assumption (Eq.~\ref{eq:beta-SFR}), shown by the red solid curve, fits better with the observations and may be superior to a constant $\beta$ assumption, as indicated by the green dashed curve and blue dotted curve. This result supports our assumed $\beta$--SFR relation in Eq. \ref{eq:beta-SFR}, which applies to all SFRs.

\subsection{IGIMFs}\label{secsub:IGIMFs}

\begin{figure}[!hbt]
    \center
    \includegraphics[width=9cm]{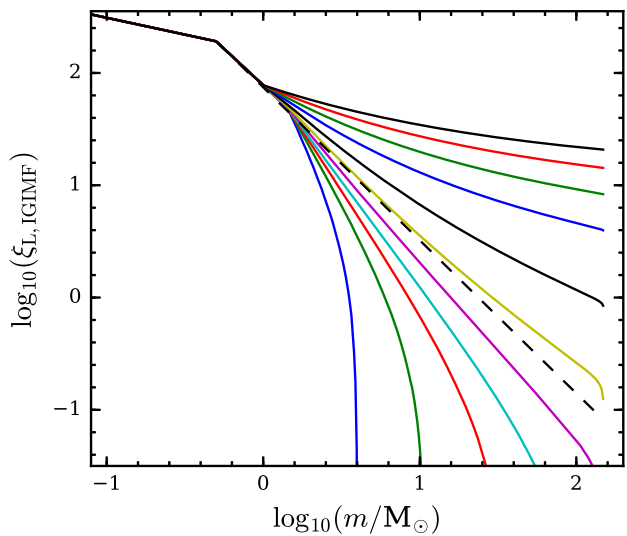}
    \caption{Logarithmic integrated galaxy-wide IMFs, $\xi_{\mathrm{L,IGIMF}}$ (Eq.~\ref{eq:xi_L_IGIMF}), for different SFRs and formed over a 10 Myr epoch. The unit of $\xi_{\mathrm{L,IGIMF}}$ is number of stars per log-mass interval. Each line is normalized to the same values at $m<1$ M$_{\odot}$. Solid curves are IGIMFs for galactic SFR=\textcolor{blue}{$10^{-5}$}, \textcolor{green}{$10^{-4}$} ... $10^{5}$ M$_{\odot}$/yr from bottom left to top right. The dashed line is the canonical IMF (Eq.~\ref{eq:xi_star}) with $\alpha_3=2.3$. The lines lower than the dashed line deviate from the canonical IMF above 1.1 M$_{\odot}$ which is the most massive stellar mass in the least-massive embedded clusters $M \approx M_{\mathrm{min}} = 5$ M$_{\odot}$. The lines on top of the dashed line deviate from the canonical IMF above 1 M$_{\odot}$ which is the lower mass limit for applying $\alpha_3$ as defined in Eq. \ref{eq:xi_star} and \ref{eq:alpha_3}. This plot is comparable with earlier IGIMF works from \cite{Weidner2013b} and \cite{Fontanot2017}.}
    \label{fig:IGIMF_SFR_}
\end{figure}

The calculated IGIMFs for different SFRs are shown in Fig.~\ref{fig:IGIMF_SFR_} and can be compared with previous IGIMF works from \cite{Weidner2013b} and \cite{Fontanot2017}. Here we define the logarithmic IMF,
\begin{equation}\label{eq:xi_L_IGIMF}
\xi_{\mathrm{L,IGIMF}}(m)=\frac{\mathrm{d}N}{\mathrm{d}\log_{10}m}=m~\ln(10)~\xi_{\mathrm{IGIMF}}(m),
\end{equation}
where $\mathrm{d}N$ is the number of stars in the logarithm mass interval $\log_{10}m$ to $\log_{10}m+\mathrm{d}\log_{10}m$ and $\xi_{\mathrm{IGIMF}}$ is defined in Eq.~\ref{eq:xi_IGIMF}. Our result differs slightly from the previous papers because of our newly applied improved optimal sampling formalism following SPK as explained in Section \ref{sec: Model-OSGIMF} and \ref{sec:discussion}.

The result shows top-heavy IMFs for high SFR galaxies and bottom-heavy IMFs for low SFR galaxies. We note that this behavior is consistent with recent observational constraints from low mass X-ray binaries \citep{Peacock2017}.

Fig.~\ref{fig:IGIMF_SFR_} also implies that galaxies with a small SFR do not form massive stars. For instance, a galaxy with a SFR$=10^{-4}$ M$_\odot$/yr will be forming no stars more massive than 10 M$_\odot$ as will be discussed in Sec.~\ref{secsub:mmax-SFR}.

\subsection{OSGIMFs}\label{secsub:OSGIMFs}

\begin{figure*}[!hbt]
    \center
    \includegraphics[width=\hsize]{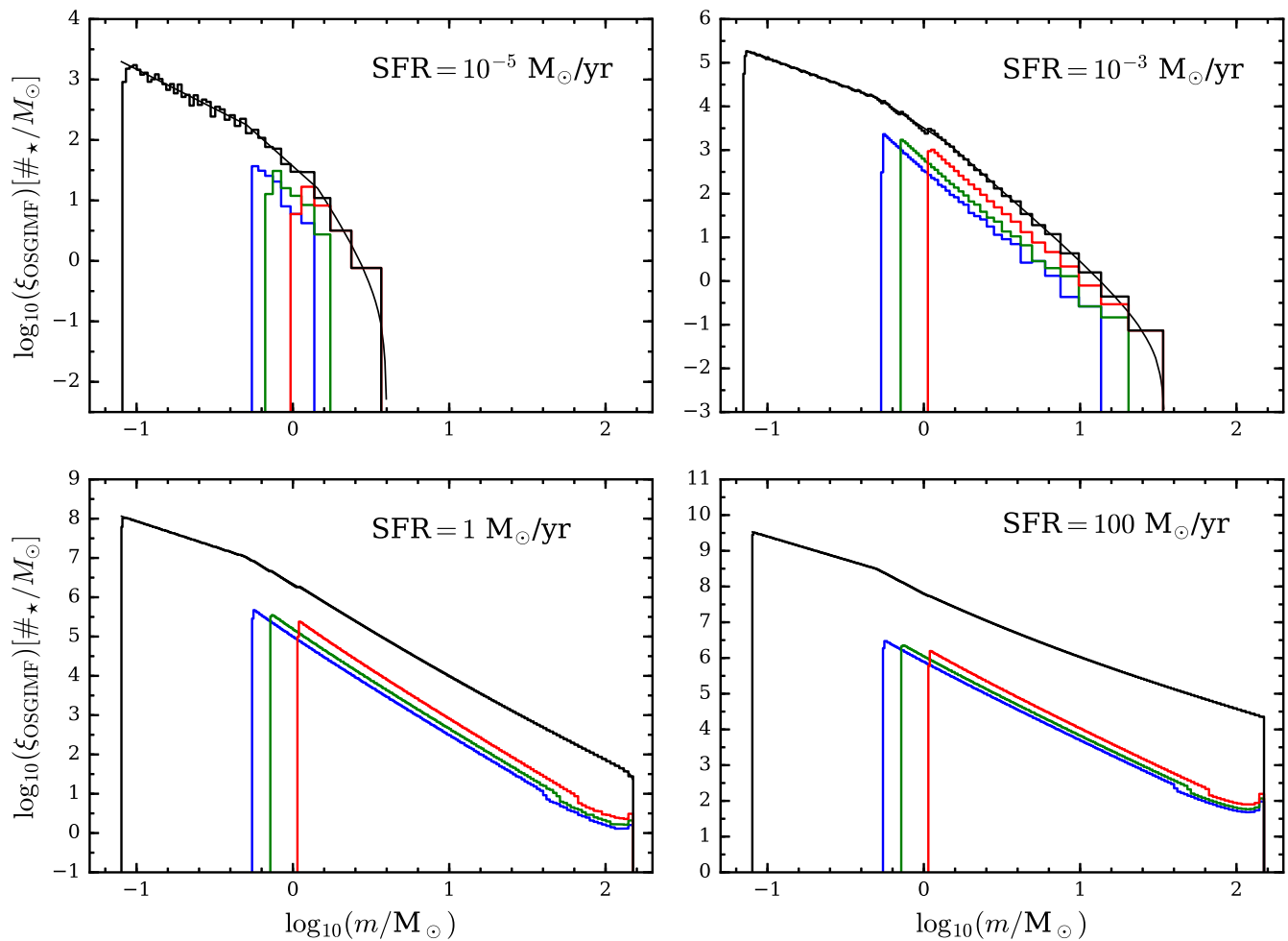}
    \caption{Optimally-sampled galaxy-wide IMFs for different SFRs. The unit of $\xi_{\mathrm{OSGIMF}}$ is number-of-stars-per-linear-mass-interval. The black histogram is the entire stellar sample in the galaxy formed in a 10 Myr epoch. The large scatter in the upper left panel and the serrated feature around $\log_{10}(m/\mathrm{M}_{\odot})=0$ for every panel is explained in the text. The \textcolor{red}{red}, the \textcolor{green}{green} and the \textcolor{blue}{blue} histograms, comprising our decomposition plot, are the OSGIMFs counting only the first, the second and the third most-massive stars of embedded clusters. Note the spoon feature at $m>10^{1.6}$ M$_\odot$ in the red, green and blue curves. The stellar mass distribution is saturated at the high mass end because of the 150 M$_{\odot}$ limit. The thin smooth curves are the IGIMFs as in Fig. \ref{fig:IGIMF_SFR_}, calculated by Eq. \ref{eq:xi_IGIMF} and normalized to give the mass in stars when integrated over the relevant stellar mass ranges.}
    \label{fig:OSGIMF}
\end{figure*}

The OSGIMFs provide the specific masses for all the individual stars in a simulated stellar system and it follows the IGIMFs tightly. The OSGIMF for different SFRs are shown in Fig.~\ref{fig:OSGIMF} with the IGIMF overlaid, where the IGIMFs are the same as in Fig. \ref{fig:IGIMF_SFR_} but here normalized to $M_{\mathrm{tot}}$ by Eq.~\ref{eq:MtotintMecl} and \ref{eq:SFR*deltat}.

The difference between the IGIMF and the OSGIMF is an additional serrated feature in the latter, arising from optimal sampling on the assumed ECMF with a sharp minimum edge at $M_{\rm min}=5$ M$_\odot$. With optimal sampling, the large number of minimum mass embedded clusters will have a similar most-massive-star mass. This is demonstrated by the decomposition plots of the OSGIMF colored in red, green, and blue, counting only the first, the second and the third massive stars of embedded clusters, respectively. i.e., the left edge of the \textcolor{red}{red} histograms in Fig.~\ref{fig:OSGIMF} is formed by the most massive stars in all the embedded clusters with mass close to $M_{\mathrm{ecl,min}}$. According to optimal sampling, those stars must have the same mass, leading to a sharp peak and the serrated feature around 1 M$_{\odot}$ that is best visible in the upper right panel and demonstrated as a derivative in Fig.~\ref{fig:OSalphaMstar} below (see caption of Fig.~\ref{fig:OSGIMF} and further discussions in Sec.~\ref{secsub:OSGIMF shape}). If a smooth changing for the ECMF lower-mass limit is assumed rather than an abrupt cutoff, then the serrated feature will be reduced.

The small drop down glitch (which makes the spoon-feature that we mentioned above) around $\log_{10}(m/M_\odot)=1.7$ in the decomposition plots is caused by the steepening feature of the $m_{\mathrm{str,max}}$--$M_{\rm ecl}$ relation in Fig.~\ref{fig:MmaxMecl}. The steepening in Fig.~\ref{fig:MmaxMecl} and the spoon-feature in Fig.~\ref{fig:OSGIMF} will disappear if a fixed $\alpha_3$ is applied, as shown in Appendix~\ref{Appendix2}.

The number of sampled stars shows a dispersion at the low mass end in the SFR $=10^{-5}$ M$_{\odot}$/yr panel, which originates from optimal sampling. Remember, we first discretely sampled clusters, then we discretely sampled stars for each cluster. So there will be coincidences of very similar stellar masses from clusters with different masses. The scatter is not Poissonian and is prominent at the low-mass end of the stellar mass function only because the bin size is sufficiently small.

\subsection{$\alpha$ plot}\label{secsub:alpha-plot}

\begin{figure}[!hbt]
    \center
    \includegraphics[width=9cm]{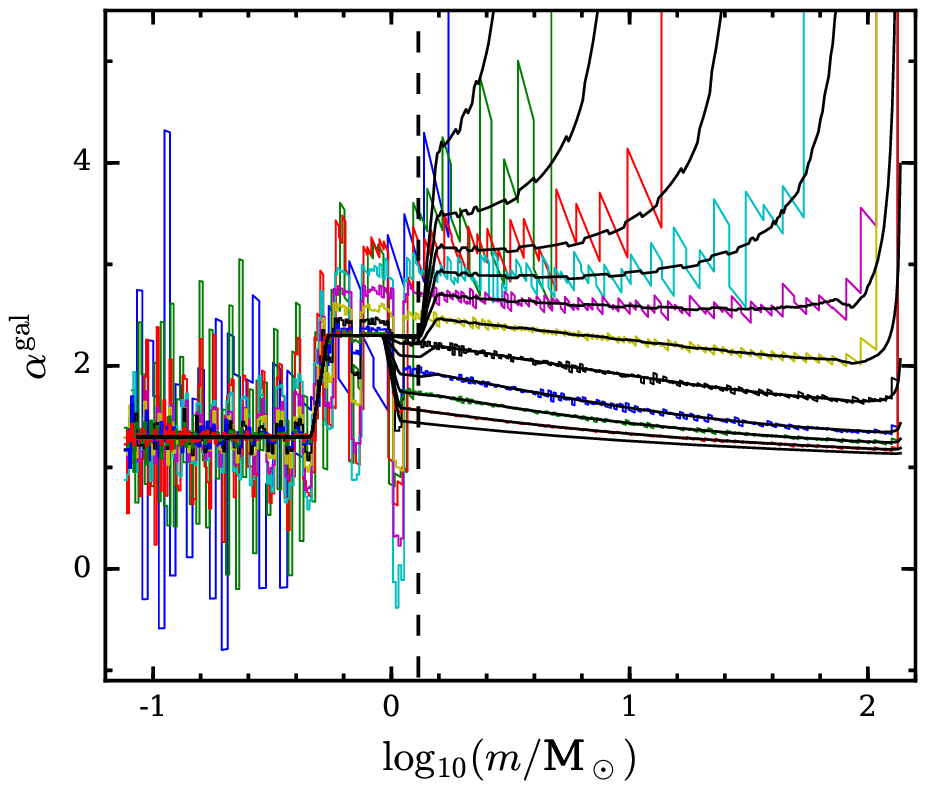}
    \caption{The galaxy-wide IMF power-law index, $\alpha^{\mathrm{gal}}$ (Eq.~\ref{eq:alpha_gal}), for the OSGIMF (colored thin lines) and the IGIMF (black lines) for SFR $=10^{-5}$, $10^{-4}$ ... $10^{5}$ M$_{\odot}$/yr from upper to lower lines. (For the SFR=$10^{5}$ M$_{\odot}$/yr case, only the IGIMF index is plotted because a star-by-star sampling costs too much computational time.) Below 1 M$_{\odot}$, the plot reproduces the canonical IMF indexes of 1.3 and 2.3 as defined in Eq. \ref{eq:xi_star}. Between 1 and 1.1 M$_{\odot}$, $\alpha^{\mathrm{gal}}$ follows the variable $\alpha_3$ defined in Eq. \ref{eq:alpha_3}. At higher masses, $\alpha^{\mathrm{gal}}$ varies with SFR. The OSGIMF index deviates from the IGIMF one around 1 M$_{\odot}$. This result may lead to a large observational scatter at intermediate and high mass ranges. The large $\alpha^{\mathrm{gal}}$ scatter for the OSGIMF at the low mass end and low SFR has been explained in the text for Fig.~\ref{fig:OSGIMF}. The vertical dashed line at $m=1.3$ M$_{\odot}$ divides the IGIMF into two parts as mentioned in Sec.~\ref{sec:discussion IGIMF shape}}
    \label{fig:OSalphaMstar}
\end{figure}

We define the incident galaxy-wide IMF power-law index or slope as:
\begin{equation}\label{eq:alpha_gal}
\alpha^{\mathrm{gal}}(m)=-\frac{\mathrm{d}(\log_{10}\xi_{\mathrm{GIMF}})}{\mathrm{d}(\log_{10} m/\mathrm{M}_{\odot})},
\end{equation}
where $\xi_{\mathrm{GIMF}}$ can be the integrated galaxy-wide IMF, $\xi_{\mathrm{IGIMF}}$, or the optimally-sampled galaxy-wide IMF, $\xi_{\mathrm{OSGIMF}}$. The alpha-plot \citep{Kroupa2001} for the OGIMF and the IGIMF for different SFRs is shown in Fig. \ref{fig:OSalphaMstar} as colored lines and black lines respectively. The figure shows a diversion at the high mass end for different SFRs and a systematic fluctuation around 1 M$_{\odot}$ which is caused by the serrated feature mentioned above.

\subsection{$\alpha_3^{\mathrm{gal}}$--SFR}\label{secsub:alpha3-SFR}

\begin{figure*}[!hbt]
    \center
    \includegraphics[width=\hsize]{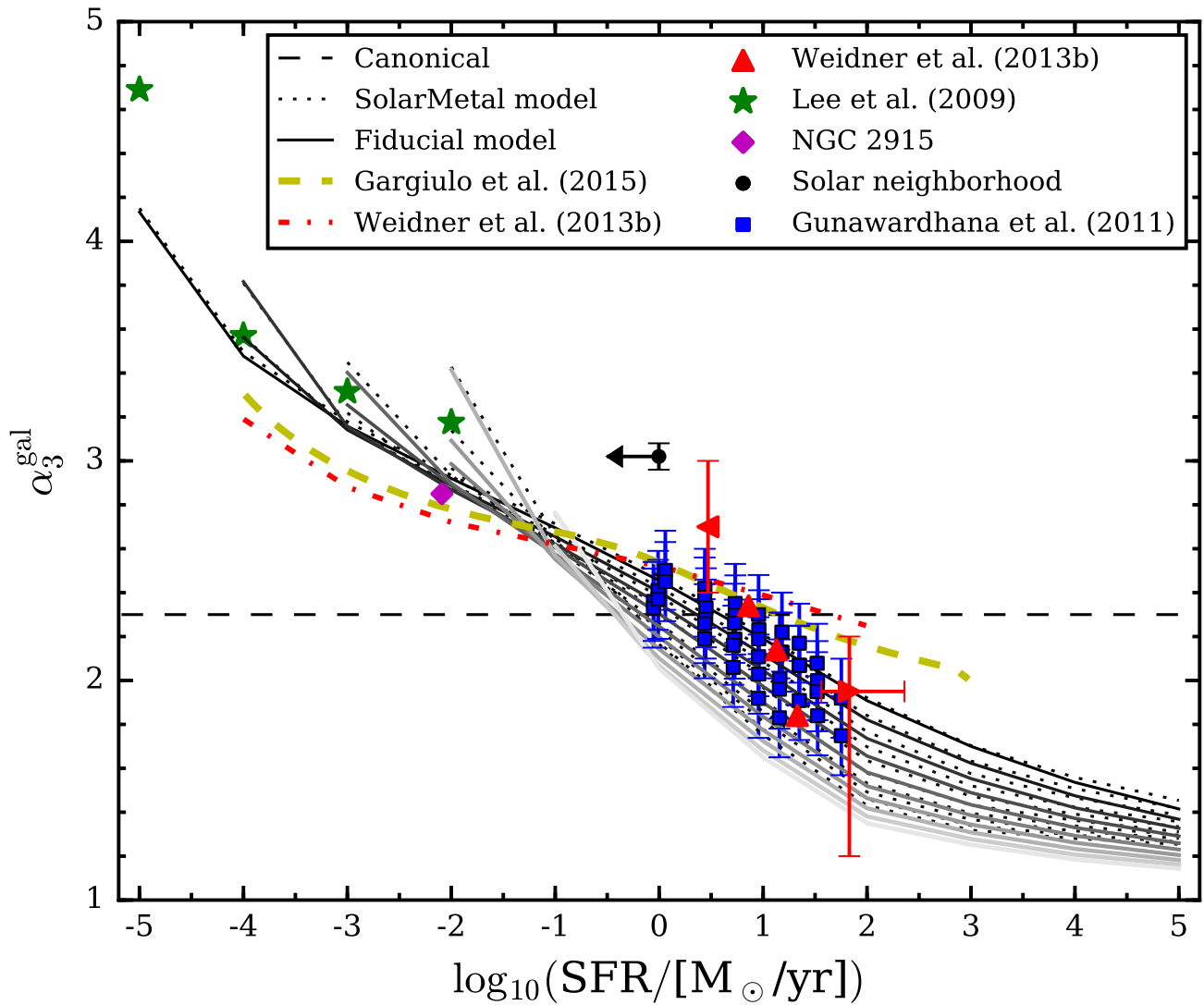}
    \caption{The observed high mass end power-law index of the galaxy-wide IMF as resulting from the here calculated IGIMF, $\alpha_3^{\mathrm{gal}}$ (i.e. $\alpha^{\mathrm{gal}}$ in Eq.~\ref{eq:alpha_gal} for $m>1$ M$_{\odot}$), for a constant SFR over $\delta t=10$ Myr in dependence of the galaxy-wide SFR. In Fig. \ref{fig:OSalphaMstar}, $\alpha_3^{\mathrm{gal}}$ values diverge for different SFRs and also vary for different $m$ at the high mass end. As at each $m$ value there exists a different $\alpha_3^{\mathrm{gal}}$--SFR relation, we plot solid lines for $\log_{10}(m/$M$_{\odot})=$0.2, 0.4, ..., 2, i.e., 1.58, 2.51, ..., 100 M$_{\odot}$ from black to gray (top to bottom) for the fiducial model and dotted lines for the corresponding SolarMetal model defined in Sec. \ref{sec:Model-IGIMF}. \textcolor{blue}{Blue squares} are data from the GAMA galaxy survey \citep{Gunawardhana2011}. \textcolor{red}{Red triangles} and the \textcolor{red}{red dash-dotted line} are data from \cite{Weidner2013b} where the left triangle is for the MW field, middle three triangles are galaxy studies, the right triangle is for the bulges of the MW and M31 and the dash-dotted line is their IGIMF model assuming $\beta=2$. A recent study has suggested that the 2 M$_{\odot}$/yr SFR for MW is overestimated \citep{Chomiuk2011} but we leave this data point the same as in \cite{Weidner2013b}. \cite{Gargiulo2015} report consistency between their IGIMF model assuming $\beta=2$ (\textcolor{orange}{thick yellow dashed line}) and the [$\alpha$/Fe] abundance ratios of elliptical galaxies. The \textcolor{purple}{purple diamond} is an individual analysis for the dwarf galaxy NGC 2915 \citep{Bruzzese2015}. \textcolor{green}{Green stars} are based on the \cite{Lee2009} 11HUGS observations of dwarf galaxies. The black circle is an observation for the solar neighborhood from \cite{Rybizki2015} with adopted MW SFR from \cite{Robitaille2010} as an upper limit of the solar neighborhood SFR because the Sun is located in an inter-arm region where the relevant SFR is significantly smaller (towards the direction indicated by the arrow, see Sec.~\ref{secsub:alpha3-SFR} for further details). The thin horizontal dashed line represents the canonical IMF index $\alpha_2=\alpha_3=2.3$.}
    \label{fig:alpha3SFR}
\end{figure*}

Fig.~\ref{fig:alpha3SFR} is the extracted $\alpha_3^{\mathrm{gal}}$--SFR relation ($\alpha_3^{\mathrm{gal}}$ is $\alpha^{\mathrm{gal}}$ in Eq.~\ref{eq:alpha_gal} for $m>1$ M$_{\odot}$) from the alpha-plot (Fig. \ref{fig:OSalphaMstar}) at different stellar masses, $m$, where the IGIMF is curved and can not be described by a single power-law. As each $m$ value gives a different $\alpha_3^{\mathrm{gal}}$--SFR relation, we plot a few lines from black to gray for stellar masses from 1.58 to 100 M$_{\odot}$. Generally we have a top-heavy IMF for large-SFR galaxies and bottom-heavy IMF for low-SFR galaxies.

Unlike the model result, the data points obtained by the respective observational studies assume a single power law IMF model\footnote{The models have a single power index either for the entire IMF or at the high stellar mass end.} The data points should therefore not be considered as IMF power-law indices probed for a specific stellar mass range and are not directly comparable with our model. For the plotted data and our model to be directly comparable, one still needs to:
1. Use the current model to simulate the initial stellar population for an entire star formation history of the galaxy instead of only a star formation epoch as performed here. 
2. Simulate the current stellar population and galaxy spectrum with star evolution models considering the age of each star. 
3. Simulate the observed spectrum considering extinction and a telescope model.
4. Use the same single power-law IMF model and all other assumptions to fit the simulated observed spectrum exactly as the observational studies did. 
As these procedures require much more investment and may only provide similar results, we use the data points in Fig.~\ref{fig:alpha3SFR} as only an indicator to show that the general trend of our model is consistent with the observational research results.

The data points of \cite{Gunawardhana2011} were extracted from the panel (a) in their figure 13. The data points of \cite{Lee2009} were extracted from the IGIMF model by \citet[their figure 4]{Pflamm-Altenburg2007}\footnote{For simplicity, the $\alpha_3$ value extracted here for Fig.~\ref{fig:alpha3SFR} is only a characteristic value for \cite{Pflamm-Altenburg2007}'s curved IGIMF shape. In principle, it should also be shown as a few lines in Fig.~\ref{fig:alpha3SFR} as for our own calculation. Also, the SFR $>2$ M$_{\odot}$/yr part of the \cite{Pflamm-Altenburg2007} result is not adopted because their stellar IMF is fixed without the $\alpha_3$ dependence assumed here in Eq. \ref{eq:alpha_3} which changes the IMF shape at higher SFRs.}. \cite{Lee2009} examined 11HUGS data of far ultraviolet non-ionizing continuum and H$\alpha$ nebular emission for a volume limited sample of about 300 nearby star-forming dwarf galaxies, finding a systematic change of the H$\alpha$ over UV flux and concluded that the IGIMF is consistent with these 11HUGS observations which indicate a systematic deficit of ionizing massive star in galaxies with low SFRs. The observational constraints from \cite{Lee2009} indicate an upwards turn in the $\alpha_3^{\mathrm{gal}}$ curves towards very small SFRs. This is, remarkably, a natural outcome of the IGMF theory, having been predicted by \cite{Pflamm-Altenburg2007,Pflamm-Altenburg2009}. The data from \cite{Weidner2013b} is extracted from their figure 1 where they assume $\beta=2$ and $\alpha_3(x)$ (our Eq.~\ref{eq:alpha_3_metal}). \cite{Gargiulo2015} checked the consistency between the IGIMF theory and [$\alpha$/Fe] abundance ratios of elliptical galaxies. They use a constant-$\beta$ and a preliminary version of the $\alpha_3$--$M_{\mathrm{ecl}}$ relation. The middle line in their figure 1 (for "$\beta=2$, $M_{\mathrm{ecl}}^{\mathrm{min}}=5$ M$_{\odot}$") is shown in our Fig.~\ref{fig:alpha3SFR} as the yellow dashed line.

Concerning the Solar neighborhood data evident in Fig.~\ref{fig:alpha3SFR} that do not appear to fit the IGIMF theory: These data points are plotted using the MW SFR $\approx 1$ M$_{\odot}$/yr estimated by \cite{Robitaille2010}. But this Galactic estimation can be higher than the local SFR in the Solar neighborhood as we are in an inter-arm region that has a relatively lower SFR than the average Galactic value as suggested by extra-galactic studies (e.g. \citealt{Seigar2002}). In addition, the local field $\alpha_3$ value can also be different from $\alpha_3^{\mathrm{gal}}$. Several past studies \citep{Kennicutt1983,Kennicutt1994,Baldry2003} indeed reported a discrepancy between the IMF of the Solar neighborhood (Scalo IMF; \citealt{Scalo1986}) and other late-type galaxies similar to the MW (Salpeter IMF; \citealt{Salpeter1955}). Variation of the local field IMF as a result of the changing local SFR are discussed by \cite{Elmegreen2006}. Thus, the local field study can only be taken as a reference for the galaxy-wide SFR--$\alpha_3^{\mathrm{gal}}$ relation of the MW.

We note that the data points shown in Fig.~\ref{fig:alpha3SFR} are not direct measurements but interpretations from flux measurements, e.g., H$\alpha$ / UV ratios. A variable IMF is not necessarily the only way to explain such observations. Differences in the treatments of processes or a special star formation history (SFH) may lead to similar results. However, the attempt of changing the SFH to account for the observations requires all the galaxies being analyzed to be in-phase in their SFH (see "coordinated bursts" in \citealt{Hoversten2008}). As the cited studies used a large number of galaxies, this scenario becomes highly unlikely. In addition, the magnitude of SFR variation required in this special SFH scenario is unrealistic as stated by \citet[their section 4.5]{Lee2009}. A comprehensive discussion can also be found at \citet[their section 6]{Gunawardhana2011}.

In addition, a recent study of the Sagittarius dwarf galaxy \citep{Hasselquist2017} performed an analysis on elemental abundances for 16 different elements on 158 red giant stars suggesting that this low SFR dwarf galaxy needed a top-light IMF.

Nevertheless, here the argument is not that the observationally suggested $\alpha_3^{\mathrm{gal}}$--SFR relation must be true, but that the apparent agreement between this suggested relation and our derived relation from the locally valid observations and our theory is remarkable. The general shape of the IGIMF model's prediction (gray lines in Fig.~\ref{fig:alpha3SFR}) follows the empirical extra-galactic constraints well and the results calculated here are comparable to previous work (\citealt{Weidner2013b,Gargiulo2015} \footnote{They both use the assumption $\beta=2$ while our $\beta$ is 2 only when $\log_{10}($SFR$)=0$ and varies from 1.5 to 2.5. This difference makes our result steeper as shown in Fig.~\ref{fig:alpha3SFR}.}). Although the comparison with the IGIMF slopes is illustrative rather than quantitative as the observational constraints are obtained by assuming the galaxy-wide IMF to be a single power-law function, this is still particularly encouraging. We emphasize as a central point that this is a remarkable outcome because the IGIMF model's prediction as shown in Fig.~\ref{fig:alpha3SFR} is sensitive to Eqs.~\ref{eq:alpha_3} and \ref{eq:beta-SFR}, i.e., the $\alpha_3$--$\rho_{\rm cl}$ and $\beta$--SFR relations, which are independently determined by different empirical data\footnote{We do simplify the empirical format of Eq.~\ref{eq:beta-SFR} in comparison with \cite{Weidner2013b} but follow the same parameters.}. The good agreement between the modeled $\alpha_3^{\mathrm{gal}}$--SFR relation and the observations is a natural result of the IGIMF theory.

\subsection{$m_{\mathrm{str,max}}$--SFR}\label{secsub:mmax-SFR}

\begin{figure}[hbt]
    \center
    \includegraphics[width=9cm]{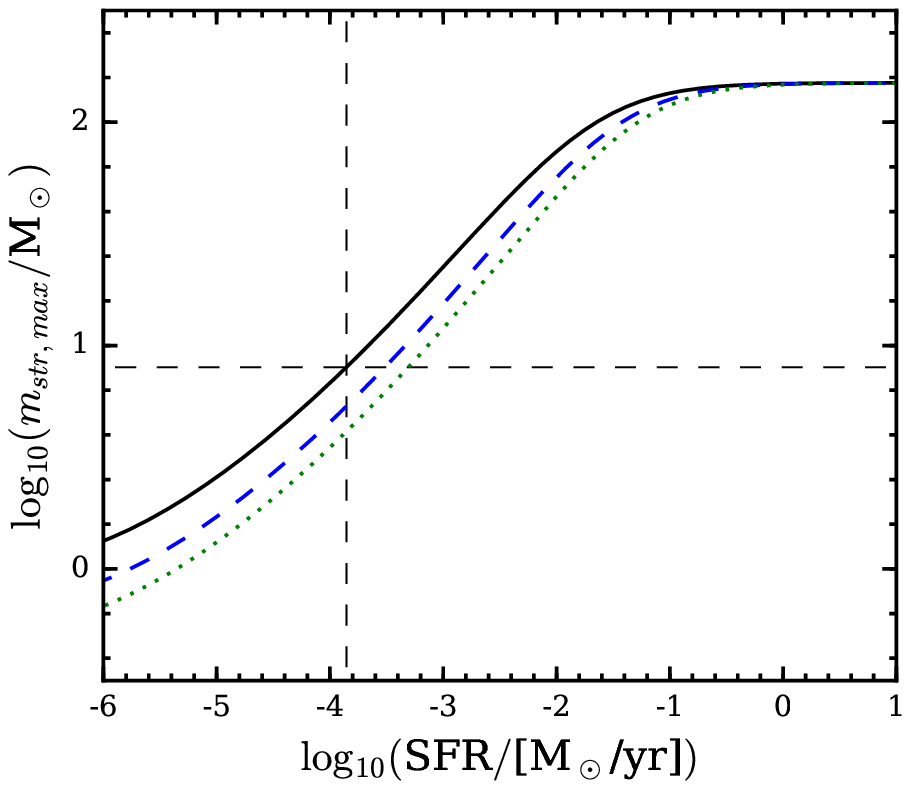}
    \caption{The optimally-sampled most-, second- and third-most massive star to be found in a galaxy as a function of the galaxy-wide SFR are shown as solid curve, \textcolor{blue}{blue dashed curve} and \textcolor{green}{green dotted curve}, respectively, resulting from a combination of Fig.~\ref{fig:MmaxMecl} and Fig.~\ref{fig:SFRMecl}. The thin horizontal dashed line indicates the mass-limit below which SNII explosions are not likely ($8\,$M$_\odot$), and  the vertical thin dashed line shows the SFR below which galaxies are not expected to host SNII events, subject to the axioms adopted in the present study. See Sec.~\ref{secsub:mmax-SFR} for more details.}
    \label{fig:MmaxSFR}
\end{figure}

Fig.~\ref{fig:MmaxSFR} shows the most-, second- and third-most massive star in a galaxy as a function of the SFR of the galaxy. It results from a combination of Fig.~\ref{fig:MmaxMecl} and Fig.~\ref{fig:SFRMecl}. 

For the here valid axioms Fig.~\ref{fig:MmaxSFR} implies that, if there were no binaries, a galaxy with a SFR $<10^{-4}$ M$_\odot$/yr would have no Type II supernova (SNII) or any other core collapse supernova events, allowing a potential observational test of the IGIMF theory as axiomatized here. 
However, since most stars are formed as binaries some of these may merge to form a massive star despite the low galaxy-wide SFR. This can result in delayed SNII events (see \citealt{Zapartas2017}).
Issues which remain uncertain are however not only that stars in some mass range and metallicity range may implode rather than explode, faking this IGIMF-induced SNII deficit. Also, the ECMF may change its character in dwarf galaxies in that massive embedded clusters may form preferentially in disks with low-shear motions \citep{Weidner2010a}, perhaps implying that below a SFR threshold the ECMF may become bimodal by containing low-mass embedded clusters and a massive one as well. In such a case the IGIMF theory remains valid since the galaxy-wide IMF remains to be the sum over all embedded clusters each of them contributing their own stellar IMF to the galaxy, but the predictions will change with the form of the ECMF. 

Nevertheless, observations of a putative deficit of SNII events in a particular complete volume of galaxies are important to test the axioms underlying the IGIMF theory. This is why Fig.~\ref{fig:MmaxSFR} is an important quantification for the particular set of axioms made in this study.

The occurrence of supernova within the IGIMF theory (but without the variation of $\alpha_3$ as implemented by Eq.~\ref{eq:alpha_3} and \ref{eq:alpha_3_metal} here) has been investigated in more detail by \cite{Weidner2005}. This is qualitatively interesting as \cite{Tsujimoto2011} noted in his chemical modeling of the Formax dSph satellite galaxy the need for a lack of stars more massive than 25 M$_\odot$.\\

In summary, the OSGIMF fulfills the currently-available observational requirements of the $m_{\mathrm{str,max}}$--$M_{\mathrm{ecl}}$, $M_{\mathrm{ecl,max}}$--SFR and $\alpha_3^{\mathrm{gal}}$--SFR relations and can be applied to galaxy evolution studies. A deficit of SNII events in galaxies with low SFRs is expected.

\section{DISCUSSION}\label{sec:discussion}

\subsection{IGIMF shape}\label{sec:discussion IGIMF shape}

The IGIMF shape in Fig.~\ref{fig:IGIMF_SFR_} shows a moderate difference from previous works \citep{Weidner2013b,Fontanot2017} because our assumptions on $\alpha_3(\rho_{\mathrm{cl}})$ and $\beta(\mathrm{SFR})$ are different, and because we use an integration upper limit rather than a maximum object mass in our Eq. \ref{eq:MtotintMecl} and \ref{eq:MeclintMstar} following SPK. For example, equation 3 and 4 of \cite{Fontanot2017}, their "$M_{\mathrm{cl}}^{\mathrm{max}}$" and "$m_{\mathrm{\star}}^{\mathrm{max}}$" were specifically defined and used as the integration upper limit in their equation 8. In the present paper, these values need to be calculated by solving our equation set \ref{eq:MtotintMecl} \& \ref{eq:1intMecl} and equation set \ref{eq:MeclintMstar} \& \ref{eq:1intMstar}. Different assumptions on $\alpha_3(\rho_{\mathrm{cl}})$ and $\beta(\mathrm{SFR})$ result in different IGIMF shapes. We list examples in Appendix \ref{Appendix}.\\

The IGIMF shape can be divided into three parts as is shown in Fig. \ref{fig:IGIMF_SFR_} \& \ref{fig:OSalphaMstar}:

(1) Below 1.3 M$_{\odot}$ and given our axioms, the IGIMF shape is identical with the stellar IMF where it has three different slopes i.e., 1.3, 2.3 and $\alpha_3$. The least massive embedded clusters, $M_{\rm ecl}\approx M_{\mathrm{min}} = 5$ M$_{\odot}$, have an optimally-sampled most massive star of $\approx 1.1$ M$_{\odot}$, so every embedded cluster can populate the mass function from 0.08 M$_{\odot}$ to a mass little larger than 1.1 M$_{\odot}$. Thus the lower mass end of the IGIMF is not influenced by cluster mass assignment. This will change if the IMF below 1 M$_{\odot}$ varies with metallicity (equation 3 in \citealt{Kroupa2002}; equation 12 in \citealt{Marks2012a}).

(2) For $m>1.1$ M$_{\odot}$ to part three below, $\alpha^{\mathrm{gal}}$ in Eq.~\ref{eq:alpha_gal}, i.e., $\alpha_3^{\mathrm{gal}}$, does not change much as shown in Fig.~\ref{fig:OSalphaMstar} and \ref{fig:alpha3SFR} especially for the high SFR cases.

When SFR $<1$ M$_{\odot}/$yr (or equivalently $\beta>2$ from Eq. \ref{eq:beta-SFR}), there is a deficit of massive clusters that are able to generate massive stars. Thus the slope of the IGIMF at the high mass end, $\alpha_3^{\mathrm{gal}}$, is largely influenced by the number of massive embedded clusters, i.e., it depends on $\beta$.

When SFR $>1$ M$_{\odot}/$yr ($\beta<2$), there are enough massive embedded clusters to fully populate the IMF. So a larger SFR resulting in more massive clusters would not change the shape of the galaxy-wide IMF anymore. Under this condition, $\alpha_3^{\mathrm{gal}}$ is only influenced by Eq.~\ref{eq:xi_star} \& \ref{eq:alpha_3}, where the IMF index $\alpha_3$ changes according to the physical condition of the molecular cloud that forms the star cluster.

(3) After the second part of the IGIMF with a relatively constant slope, the IGIMF bends down rapidly and behaves like a Schechter-type function for SFR $<10^{-3}$ M$_{\odot}$/yr at the high mass end.

\subsection{OSGIMF shape}\label{secsub:OSGIMF shape}

The OSGIMF mimics the IGIMF but shows additional features. It differs from the IGIMF because the optimally-sampled result is a discrete function under binning. If all the clusters in the model would have the same mass, optimal sampling will generate the same list of stars for all these clusters. This leads to gaps between the sampled stellar masses and peaks at the sampled masses.

In reality, clusters with different masses can smoothen this effect. With the currently used ECMF (\ref{axiom: ECMF} in Sec.\ref{sec:Model-IGIMF}), there exists a large fraction of clusters with mass $M_{\mathrm{ecl}} \approx$ 5 or 6 M$_{\odot}$. The serrated features are the legacy of these clusters. Only when the ECMF slope $\beta<2.1$, there can be enough high mass clusters all with different masses to smoothen out the serrated feature. If a smoothly varying probability distribution of cluster masses is assumed at the mass boundaries rather than the sharp-edged ECMF in Eq.~\ref{eq:xi_ecl}, there would also be no serrated features.

The above discussion only applies to the perfect deterministic scenario. Although astrophysical nature shows evidence to be closer to deterministic, i.e., a high level of self-regulation, it still possesses some level of randomness due to additional parameters not known to the theory. For instance, the angular momenta of clouds when star clusters form are not likely to be the same. This can result in a different stellar list for the embedded star clusters with the same mass and smoothen the final result.

For the present purpose, here we point out that if nature is somewhere close to the optimal sampling of stellar masses, there will be physical consequences. In particular the SFRs, stellar-mass buildup times and gas-depletion time scales of galaxies are affected significantly \citep{Pflamm-Altenburg2007,Pflamm-Altenburg2008,Pflamm-Altenburg2009ApJ}. Different sampling methods with the same cluster scale IMF actually lead to different physical results. If the serrated feature is detected, it will strongly support optimal sampling and give direct information on the ECMF.

\section{CONCLUSIONS}\label{sec:CONCLUSIONS}

We have updated the $m_{\mathrm{str,max}}$--$M_{\mathrm{ecl}}$ and $M_{\mathrm{ecl,max}}$--SFR relation with newest observations. Comparison of the optimal sampling and random sampling predictions, indicates that the optimal sampling scenario is favored.

The $\alpha_3^{\mathrm{gal}}$--SFR relation as a natural outcome of the IGIMF theory is also updated. It agrees with the observationally suggested trend.

For the first time, optimal sampling is introduced in the context of the IGIMF theory to generate a list of definite masses of stars in a galaxy. As a side effect, this also makes the sampling of stellar masses in a very massive galaxy computationally efficient\footnote{Unlike random sampling, the optimal sampling follows a deterministic equation, giving a quick method to calculate the number of stars in a given mass range without sampling each star one by one.}.

We test and apply the improved optimal sampling and normalization method to further constrain the IGIMF theory. A prediction of the IGIMF model as defined by the here adopted axioms is that dwarf galaxies with SFR $<10^{-4}$ M$_\odot$/yr should host no SNII events in the absence of binary stars.

The OSGIMF largely agrees with previous IGIMF studies and shows a special serrated structure that previous integration calculations did not contain. This structure, if existing, will be strong evidence for a deterministic and self-regulated star formation process.

Finally, we provide the publicly available Python computer module GalIMF for general use.

\medskip

\bibliography{library}

\begin{thebibliography}{95}
\expandafter\ifx\csname natexlab\endcsname\relax\def\natexlab#1{#1}\fi

\bibitem[{{Andrews} {et~al.}(2013){Andrews}, {Calzetti}, {Chandar}, {Lee},
  {Elmegreen}, {Kennicutt}, {Whitmore}, {Kissel}, {da Silva}, {Krumholz},
  {O'Connell}, {Dopita}, {Frogel}, \& {Kim}}]{Andrews2013}
{Andrews}, J.~E., {Calzetti}, D., {Chandar}, R., {et~al.} 2013, \apj, 767, 51

\bibitem[{{Baldry} \& {Glazebrook}(2003)}]{Baldry2003}
{Baldry}, I.~K. \& {Glazebrook}, K. 2003, \apj, 593, 258

\bibitem[{Banerjee \& Kroupa(2015)}]{Banerjee2015}
Banerjee, S. \& Kroupa, P. 2015 [\eprint[arXiv]{1512.03074}]

\bibitem[{{Banerjee} {et~al.}(2012{\natexlab{a}}){Banerjee}, {Kroupa}, \&
  {Oh}}]{Banerjee2012a}
{Banerjee}, S., {Kroupa}, P., \& {Oh}, S. 2012{\natexlab{a}}, \apj, 746, 15

\bibitem[{{Banerjee} {et~al.}(2012{\natexlab{b}}){Banerjee}, {Kroupa}, \&
  {Oh}}]{Banerjee2012b}
{Banerjee}, S., {Kroupa}, P., \& {Oh}, S. 2012{\natexlab{b}}, \mnras, 426, 1416

\bibitem[{{Bastian} {et~al.}(2010){Bastian}, {Covey}, \& {Meyer}}]{Bastian2010}
{Bastian}, N., {Covey}, K.~R., \& {Meyer}, M.~R. 2010, \araa, 48, 339

\bibitem[{{Bruzzese} {et~al.}(2015){Bruzzese}, {Meurer}, {Lagos}, {Elson},
  {Werk}, {Blakeslee}, \& {Ford}}]{Bruzzese2015}
{Bruzzese}, S.~M., {Meurer}, G.~R., {Lagos}, C.~D.~P., {et~al.} 2015, \mnras,
  447, 618

\bibitem[{{Chabrier}(2003)}]{Chabrier2003}
{Chabrier}, G. 2003, \pasp, 115, 763

\bibitem[{{Chomiuk} \& {Povich}(2011)}]{Chomiuk2011}
{Chomiuk}, L. \& {Povich}, M.~S. 2011, \aj, 142, 197

\bibitem[{{da Silva} {et~al.}(2012){da Silva}, {Fumagalli}, \&
  {Krumholz}}]{daSilva2012}
{da Silva}, R.~L., {Fumagalli}, M., \& {Krumholz}, M. 2012, \apj, 745, 145

\bibitem[{{Dabringhausen} {et~al.}(2008){Dabringhausen}, {Hilker}, \&
  {Kroupa}}]{Dabringhausen2008}
{Dabringhausen}, J., {Hilker}, M., \& {Kroupa}, P. 2008, \mnras, 386, 864

\bibitem[{{Dabringhausen} {et~al.}(2009){Dabringhausen}, {Kroupa}, \&
  {Baumgardt}}]{Dabringhausen2009}
{Dabringhausen}, J., {Kroupa}, P., \& {Baumgardt}, H. 2009, \mnras, 394, 1529

\bibitem[{{Dabringhausen} {et~al.}(2012){Dabringhausen}, {Kroupa},
  {Pflamm-Altenburg}, \& {Mieske}}]{Dabringhausen2012}
{Dabringhausen}, J., {Kroupa}, P., {Pflamm-Altenburg}, J., \& {Mieske}, S.
  2012, \apj, 747, 72

\bibitem[{{Dib} {et~al.}(2007){Dib}, {Kim}, \& {Shadmehri}}]{Dib2007}
{Dib}, S., {Kim}, J., \& {Shadmehri}, M. 2007, \mnras, 381, L40

\bibitem[{{Dib} {et~al.}(2017){Dib}, {Schmeja}, \& {Hony}}]{Dib2017}
{Dib}, S., {Schmeja}, S., \& {Hony}, S. 2017, \mnras, 464, 1738

\bibitem[{{Disney} {et~al.}(2008){Disney}, {Romano}, {Garcia-Appadoo}, {West},
  {Dalcanton}, \& {Cortese}}]{Disney2008}
{Disney}, M.~J., {Romano}, J.~D., {Garcia-Appadoo}, D.~A., {et~al.} 2008, \nat,
  455, 1082

\bibitem[{{Egusa} {et~al.}(2009){Egusa}, {Kohno}, {Sofue}, {Nakanishi}, \&
  {Komugi}}]{Egusa2009}
{Egusa}, F., {Kohno}, K., {Sofue}, Y., {Nakanishi}, H., \& {Komugi}, S. 2009,
  \apj, 697, 1870

\bibitem[{{Egusa} {et~al.}(2004){Egusa}, {Sofue}, \& {Nakanishi}}]{Egusa2004}
{Egusa}, F., {Sofue}, Y., \& {Nakanishi}, H. 2004, \pasj, 56, L45

\bibitem[{{Elmegreen} \& {Scalo}(2006)}]{Elmegreen2006}
{Elmegreen}, B.~G. \& {Scalo}, J. 2006, \apj, 636, 149

\bibitem[{{Elmegreen} \& {Shadmehri}(2003)}]{Elmegreen2003}
{Elmegreen}, B.~G. \& {Shadmehri}, M. 2003, \mnras, 338, 817

\bibitem[{{Figer}(2005)}]{Figer2005}
{Figer}, D.~F. 2005, \nat, 434, 192

\bibitem[{{Fontanot} {et~al.}(2017){Fontanot}, {De Lucia}, {Hirschmann},
  {Bruzual}, {Charlot}, \& {Zibetti}}]{Fontanot2017}
{Fontanot}, F., {De Lucia}, G., {Hirschmann}, M., {et~al.} 2017, \mnras, 464,
  3812

\bibitem[{{Fukui} \& {Kawamura}(2010)}]{Fukui2010}
{Fukui}, Y. \& {Kawamura}, A. 2010, \araa, 48, 547

\bibitem[{{Fumagalli} {et~al.}(2011){Fumagalli}, {da Silva}, \&
  {Krumholz}}]{Fumagalli2011}
{Fumagalli}, M., {da Silva}, R.~L., \& {Krumholz}, M.~R. 2011, \apjl, 741, L26

\bibitem[{{Gargiulo} {et~al.}(2015){Gargiulo}, {Cora}, {Padilla}, {Mu{\~n}oz
  Arancibia}, {Ruiz}, {Orsi}, {Tecce}, {Weidner}, \& {Bruzual}}]{Gargiulo2015}
{Gargiulo}, I.~D., {Cora}, S.~A., {Padilla}, N.~D., {et~al.} 2015, \mnras, 446,
  3820

\bibitem[{{Gunawardhana} {et~al.}(2011){Gunawardhana}, {Hopkins}, {Sharp},
  {Brough}, {Taylor}, {Bland-Hawthorn}, {Maraston}, {Tuffs}, {Popescu},
  {Wijesinghe}, {Jones}, {Croom}, {Sadler}, {Wilkins}, {Driver}, {Liske},
  {Norberg}, {Baldry}, {Bamford}, {Loveday}, {Peacock}, {Robotham}, {Zucker},
  {Parker}, {Conselice}, {Cameron}, {Frenk}, {Hill}, {Kelvin}, {Kuijken},
  {Madore}, {Nichol}, {Parkinson}, {Pimbblet}, {Prescott}, {Sutherland},
  {Thomas}, \& {van Kampen}}]{Gunawardhana2011}
{Gunawardhana}, M.~L.~P., {Hopkins}, A.~M., {Sharp}, R.~G., {et~al.} 2011,
  \mnras, 415, 1647

\bibitem[{{Gvaramadze} {et~al.}(2012){Gvaramadze}, {Weidner}, {Kroupa}, \&
  {Pflamm-Altenburg}}]{Gvaramadze2012}
{Gvaramadze}, V.~V., {Weidner}, C., {Kroupa}, P., \& {Pflamm-Altenburg}, J.
  2012, \mnras, 424, 3037

\bibitem[{{Haghi} {et~al.}(2017){Haghi}, {Khalaj}, {Hasani Zonoozi}, \&
  {Kroupa}}]{Haghi2017}
{Haghi}, H., {Khalaj}, P., {Hasani Zonoozi}, A., \& {Kroupa}, P. 2017, \apj,
  839, 60

\bibitem[{{Hasselquist} {et~al.}(2017){Hasselquist}, {Shetrone}, {Smith},
  {Holtzman}, {McWilliam}, {Fern{\'a}ndez-Trincado}, {Beers}, {Majewski},
  {Nidever}, {Tang}, {Tissera}, {Fern{\'a}ndez Alvar}, {Allende Prieto},
  {Almeida}, {Anguiano}, {Battaglia}, {Carigi}, {Delgado Inglada},
  {Frinchaboy}, {Garc{\'{\i}}a-Hern{\'a}ndez}, {Geisler}, {Minniti}, {Placco},
  {Schultheis}, {Sobeck}, \& {Villanova}}]{Hasselquist2017}
{Hasselquist}, S., {Shetrone}, M., {Smith}, V., {et~al.} 2017, ArXiv e-prints
  [\eprint[arXiv]{1707.03456}]

\bibitem[{{Hopkins}(2013)}]{Hopkins2013}
{Hopkins}, P.~F. 2013, \mnras, 433, 170

\bibitem[{{Hoversten} \& {Glazebrook}(2008)}]{Hoversten2008}
{Hoversten}, E.~A. \& {Glazebrook}, K. 2008, \apj, 675, 163

\bibitem[{{Hsu} {et~al.}(2012){Hsu}, {Hartmann}, {Allen}, {Hern{\'a}ndez},
  {Megeath}, {Mosby}, {Tobin}, \& {Espaillat}}]{Hsu2012}
{Hsu}, W.-H., {Hartmann}, L., {Allen}, L., {et~al.} 2012, \apj, 752, 59

\bibitem[{{Hsu} {et~al.}(2013){Hsu}, {Hartmann}, {Allen}, {Hern{\'a}ndez},
  {Megeath}, {Tobin}, \& {Ingleby}}]{Hsu2013}
{Hsu}, W.-H., {Hartmann}, L., {Allen}, L., {et~al.} 2013, \apj, 764, 114

\bibitem[{{Kennicutt}(1983)}]{Kennicutt1983}
{Kennicutt}, Jr., R.~C. 1983, \apj, 272, 54

\bibitem[{{Kennicutt} {et~al.}(1994){Kennicutt}, {Tamblyn}, \&
  {Congdon}}]{Kennicutt1994}
{Kennicutt}, Jr., R.~C., {Tamblyn}, P., \& {Congdon}, C.~E. 1994, \apj, 435, 22

\bibitem[{{Kirk} \& {Myers}(2011)}]{Kirk2011}
{Kirk}, H. \& {Myers}, P.~C. 2011, \apj, 727, 64

\bibitem[{{Kirk} \& {Myers}(2012)}]{Kirk2012}
{Kirk}, H. \& {Myers}, P.~C. 2012, \apj, 745, 131

\bibitem[{{Koen}(2006)}]{Koen2006}
{Koen}, C. 2006, \mnras, 365, 590

\bibitem[{{Kroupa}(1995{\natexlab{a}})}]{Kroupa1995a}
{Kroupa}, P. 1995{\natexlab{a}}, \mnras, 277 [\eprint{astro-ph/9508117}]

\bibitem[{{Kroupa}(1995{\natexlab{b}})}]{Kroupa1995b}
{Kroupa}, P. 1995{\natexlab{b}}, \mnras, 277 [\eprint{astro-ph/9508084}]

\bibitem[{{Kroupa}(2001)}]{Kroupa2001}
{Kroupa}, P. 2001, \mnras, 322, 231

\bibitem[{{Kroupa}(2002)}]{Kroupa2002}
{Kroupa}, P. 2002, Science, 295, 82

\bibitem[{{Kroupa}(2005)}]{Kroupa2005}
{Kroupa}, P. 2005, in ESA Special Publication, Vol. 576, The Three-Dimensional
  Universe with Gaia, ed. C.~{Turon}, K.~S. {O'Flaherty}, \& M.~A.~C.
  {Perryman}, 629

\bibitem[{{Kroupa}(2015)}]{Kroupa2015}
{Kroupa}, P. 2015, Canadian Journal of Physics, 93, 169

\bibitem[{{Kroupa} \& {Bouvier}(2003)}]{Kroupa2003a}
{Kroupa}, P. \& {Bouvier}, J. 2003, \mnras, 346, 343

\bibitem[{{Kroupa} \& {Weidner}(2003)}]{Kroupa2003}
{Kroupa}, P. \& {Weidner}, C. 2003, \apj, 598, 1076

\bibitem[{{Kroupa} {et~al.}(2013){Kroupa}, {Weidner}, {Pflamm-Altenburg},
  {Thies}, {Dabringhausen}, {Marks}, \& {Maschberger}}]{Kroupa2013}
{Kroupa}, P., {Weidner}, C., {Pflamm-Altenburg}, J., {et~al.} 2013, {The
  Stellar and Sub-Stellar Initial Mass Function of Simple and Composite
  Populations}, ed. T.~D. {Oswalt} \& G.~{Gilmore}, 115

\bibitem[{{Lada} \& {Lada}(2003)}]{Lada2003}
{Lada}, C.~J. \& {Lada}, E.~A. 2003, \araa, 41, 57

\bibitem[{{Lee} {et~al.}(2009){Lee}, {Gil de Paz}, {Tremonti}, {Kennicutt},
  {Salim}, {Bothwell}, {Calzetti}, {Dalcanton}, {Dale}, {Engelbracht}, {Funes},
  {Johnson}, {Sakai}, {Skillman}, {van Zee}, {Walter}, \& {Weisz}}]{Lee2009}
{Lee}, J.~C., {Gil de Paz}, A., {Tremonti}, C., {et~al.} 2009, \apj, 706, 599

\bibitem[{{Leisawitz}(1989)}]{Leisawitz1989}
{Leisawitz}, D. 1989, in \baas, Vol.~21, Bulletin of the American Astronomical
  Society, 1189

\bibitem[{{Lieberz} \& {Kroupa}(2017)}]{Lieberz2017}
{Lieberz}, P. \& {Kroupa}, P. 2017, \mnras, 465, 3775

\bibitem[{{Ma{\'{\i}}z Apell{\'a}niz} {et~al.}(2007){Ma{\'{\i}}z
  Apell{\'a}niz}, {Walborn}, {Morrell}, {Niemela}, \& {Nelan}}]{Maiz2007}
{Ma{\'{\i}}z Apell{\'a}niz}, J., {Walborn}, N.~R., {Morrell}, N.~I., {Niemela},
  V.~S., \& {Nelan}, E.~P. 2007, \apj, 660, 1480

\bibitem[{{Marks} \& {Kroupa}(2012)}]{Marks2012}
{Marks}, M. \& {Kroupa}, P. 2012, \aap, 543, A8

\bibitem[{{Marks} {et~al.}(2012){Marks}, {Kroupa}, {Dabringhausen}, \&
  {Pawlowski}}]{Marks2012a}
{Marks}, M., {Kroupa}, P., {Dabringhausen}, J., \& {Pawlowski}, M.~S. 2012,
  \mnras, 422, 2246

\bibitem[{{Matteucci} \& {Brocato}(1990)}]{Matteucci1990}
{Matteucci}, F. \& {Brocato}, E. 1990, \apj, 365, 539

\bibitem[{{Megeath} {et~al.}(2016){Megeath}, {Gutermuth}, {Muzerolle},
  {Kryukova}, {Hora}, {Allen}, {Flaherty}, {Hartmann}, {Myers}, {Pipher},
  {Stauffer}, {Young}, \& {Fazio}}]{Megeath2016}
{Megeath}, S.~T., {Gutermuth}, R., {Muzerolle}, J., {et~al.} 2016, \aj, 151, 5

\bibitem[{{Meidt} {et~al.}(2015){Meidt}, {Hughes}, {Dobbs}, {Pety}, {Thompson},
  {Garc{\'{\i}}a-Burillo}, {Leroy}, {Schinnerer}, {Colombo}, {Querejeta},
  {Kramer}, {Schuster}, \& {Dumas}}]{Meidt2015}
{Meidt}, S.~E., {Hughes}, A., {Dobbs}, C.~L., {et~al.} 2015, \apj, 806, 72

\bibitem[{{Meurer} {et~al.}(2009){Meurer}, {Wong}, {Kim}, {Hanish}, {Heckman},
  {Werk}, {Bland-Hawthorn}, {Dopita}, {Zwaan}, {Koribalski}, {Seibert},
  {Thilker}, {Ferguson}, {Webster}, {Putman}, {Knezek}, {Doyle}, {Drinkwater},
  {Hoopes}, {Kilborn}, {Meyer}, {Ryan-Weber}, {Smith}, \&
  {Staveley-Smith}}]{Meurer2009}
{Meurer}, G.~R., {Wong}, O.~I., {Kim}, J.~H., {et~al.} 2009, \apj, 695, 765

\bibitem[{{Oey} \& {Clarke}(2005)}]{Oey2005}
{Oey}, M.~S. \& {Clarke}, C.~J. 2005, \apjl, 620, L43

\bibitem[{{Oh} \& {Kroupa}(2012)}]{Oh2012}
{Oh}, S. \& {Kroupa}, P. 2012, \mnras, 424, 65

\bibitem[{{Oh} \& {Kroupa}(2017)}]{Oh2017}
{Oh}, S. \& {Kroupa}, P. 2017

\bibitem[{{Peacock} {et~al.}(2017){Peacock}, {Zepf}, {Kundu}, {Maccarone},
  {Lehmer}, {Maraston}, {Gonzalez}, {Eufrasio}, \& {Coulter}}]{Peacock2017}
{Peacock}, M.~B., {Zepf}, S.~E., {Kundu}, A., {et~al.} 2017, \apj, 841, 28

\bibitem[{{Pflamm-Altenburg} {et~al.}(2013){Pflamm-Altenburg},
  {Gonz{\'a}lez-L{\'o}pezlira}, \& {Kroupa}}]{Pflamm-Altenburg2013}
{Pflamm-Altenburg}, J., {Gonz{\'a}lez-L{\'o}pezlira}, R.~A., \& {Kroupa}, P.
  2013, \mnras, 435, 2604

\bibitem[{{Pflamm-Altenburg} \& {Kroupa}(2008)}]{Pflamm-Altenburg2008}
{Pflamm-Altenburg}, J. \& {Kroupa}, P. 2008, \nat, 455, 641

\bibitem[{{Pflamm-Altenburg} \& {Kroupa}(2009)}]{Pflamm-Altenburg2009ApJ}
{Pflamm-Altenburg}, J. \& {Kroupa}, P. 2009, \apj, 706, 516

\bibitem[{{Pflamm-Altenburg} \& {Kroupa}(2010)}]{Pflamm-Altenburg2010}
{Pflamm-Altenburg}, J. \& {Kroupa}, P. 2010, \mnras, 404, 1564

\bibitem[{{Pflamm-Altenburg} {et~al.}(2007){Pflamm-Altenburg}, {Weidner}, \&
  {Kroupa}}]{Pflamm-Altenburg2007}
{Pflamm-Altenburg}, J., {Weidner}, C., \& {Kroupa}, P. 2007, \apj, 671, 1550

\bibitem[{{Pflamm-Altenburg} {et~al.}(2009){Pflamm-Altenburg}, {Weidner}, \&
  {Kroupa}}]{Pflamm-Altenburg2009}
{Pflamm-Altenburg}, J., {Weidner}, C., \& {Kroupa}, P. 2009, \mnras, 395, 394

\bibitem[{{Ram{\'{\i}}rez Alegr{\'{\i}}a} {et~al.}(2016){Ram{\'{\i}}rez
  Alegr{\'{\i}}a}, {Borissova}, {Chen{\'e}}, {Bonatto}, {Kurtev}, {Amigo},
  {Kuhn}, {Gromadzki}, \& {Carballo-Bello}}]{RamirezAlegria2016}
{Ram{\'{\i}}rez Alegr{\'{\i}}a}, S., {Borissova}, J., {Chen{\'e}}, A.-N.,
  {et~al.} 2016, \aap, 588, A40

\bibitem[{{Randriamanakoto} {et~al.}(2013){Randriamanakoto}, {Escala},
  {V{\"a}is{\"a}nen}, {Kankare}, {Kotilainen}, {Mattila}, \&
  {Ryder}}]{Randriamanakoto2013}
{Randriamanakoto}, Z., {Escala}, A., {V{\"a}is{\"a}nen}, P., {et~al.} 2013,
  \apjl, 775, L38

\bibitem[{{Recchi} \& {Kroupa}(2015)}]{Recchi2015}
{Recchi}, S. \& {Kroupa}, P. 2015, \mnras, 446, 4168

\bibitem[{{Renaud} {et~al.}(2016){Renaud}, {Famaey}, \& {Kroupa}}]{Renaud2016}
{Renaud}, F., {Famaey}, B., \& {Kroupa}, P. 2016, \mnras, 463, 3637

\bibitem[{{Robitaille} \& {Whitney}(2010)}]{Robitaille2010}
{Robitaille}, T.~P. \& {Whitney}, B.~A. 2010, \apjl, 710, L11

\bibitem[{{Rybizki} \& {Just}(2015)}]{Rybizki2015}
{Rybizki}, J. \& {Just}, A. 2015, \mnras, 447, 3880

\bibitem[{{Salpeter}(1955)}]{Salpeter1955}
{Salpeter}, E.~E. 1955, \apj, 121, 161

\bibitem[{{Scalo}(1986)}]{Scalo1986}
{Scalo}, J.~M. 1986, in IAU Symposium, Vol. 116, Luminous Stars and
  Associations in Galaxies, ed. C.~W.~H. {De Loore}, A.~J. {Willis}, \&
  P.~{Laskarides}, 451--466

\bibitem[{{Schulz} {et~al.}(2015){Schulz}, {Pflamm-Altenburg}, \&
  {Kroupa}}]{Schulz2015}
{Schulz}, C., {Pflamm-Altenburg}, J., \& {Kroupa}, P. 2015, \aap, 582, A93

\bibitem[{{Seigar} \& {James}(2002)}]{Seigar2002}
{Seigar}, M.~S. \& {James}, P.~A. 2002, \mnras, 337, 1113

\bibitem[{{Shadmehri}(2004)}]{Shadmehri2004}
{Shadmehri}, M. 2004, \mnras, 354, 375

\bibitem[{{Stephens} {et~al.}(2017){Stephens}, {Gouliermis}, {Looney},
  {Gruendl}, {Chu}, {Weisz}, {Seale}, {Chen}, {Wong}, {Hughes}, {Pineda},
  {Ott}, \& {Muller}}]{Stephens2017}
{Stephens}, I.~W., {Gouliermis}, D., {Looney}, L.~W., {et~al.} 2017, \apj, 834,
  94

\bibitem[{{Telford} {et~al.}(2016){Telford}, {Dalcanton}, {Skillman}, \&
  {Conroy}}]{Telford2016}
{Telford}, O.~G., {Dalcanton}, J.~J., {Skillman}, E.~D., \& {Conroy}, C. 2016,
  \apj, 827, 35

\bibitem[{{Thies} {et~al.}(2015){Thies}, {Pflamm-Altenburg}, {Kroupa}, \&
  {Marks}}]{Thies2015}
{Thies}, I., {Pflamm-Altenburg}, J., {Kroupa}, P., \& {Marks}, M. 2015, \apj,
  800, 72

\bibitem[{{Tsujimoto}(2011)}]{Tsujimoto2011}
{Tsujimoto}, T. 2011, \apj, 736, 113

\bibitem[{{Vazdekis} {et~al.}(2003){Vazdekis}, {Cenarro}, {Gorgas}, {Cardiel},
  \& {Peletier}}]{Vazdekis2003}
{Vazdekis}, A., {Cenarro}, A.~J., {Gorgas}, J., {Cardiel}, N., \& {Peletier},
  R.~F. 2003, \mnras, 340, 1317

\bibitem[{{Weidner} {et~al.}(2010){Weidner}, {Bonnell}, \&
  {Zinnecker}}]{Weidner2010a}
{Weidner}, C., {Bonnell}, I.~A., \& {Zinnecker}, H. 2010, \apj, 724, 1503

\bibitem[{{Weidner} \& {Kroupa}(2004)}]{Weidner2004a}
{Weidner}, C. \& {Kroupa}, P. 2004, \mnras, 348, 187

\bibitem[{{Weidner} \& {Kroupa}(2005)}]{Weidner2005}
{Weidner}, C. \& {Kroupa}, P. 2005, \apj, 625, 754

\bibitem[{{Weidner} \& {Kroupa}(2006)}]{Weidner2006}
{Weidner}, C. \& {Kroupa}, P. 2006, \mnras, 365, 1333

\bibitem[{{Weidner} {et~al.}(2004){Weidner}, {Kroupa}, \&
  {Larsen}}]{Weidner2004}
{Weidner}, C., {Kroupa}, P., \& {Larsen}, S.~S. 2004, \mnras, 350, 1503

\bibitem[{{Weidner} {et~al.}(2013{\natexlab{a}}){Weidner}, {Kroupa}, \&
  {Pflamm-Altenburg}}]{Weidner2013a}
{Weidner}, C., {Kroupa}, P., \& {Pflamm-Altenburg}, J. 2013{\natexlab{a}},
  \mnras, 434, 84

\bibitem[{{Weidner} {et~al.}(2014){Weidner}, {Kroupa}, \&
  {Pflamm-Altenburg}}]{Weidner2014}
{Weidner}, C., {Kroupa}, P., \& {Pflamm-Altenburg}, J. 2014, \mnras, 441, 3348

\bibitem[{{Weidner} {et~al.}(2013{\natexlab{b}}){Weidner}, {Kroupa},
  {Pflamm-Altenburg}, \& {Vazdekis}}]{Weidner2013b}
{Weidner}, C., {Kroupa}, P., {Pflamm-Altenburg}, J., \& {Vazdekis}, A.
  2013{\natexlab{b}}, \mnras, 436, 3309

\bibitem[{{Yan} {et~al.}(2017){Yan}, {Jerabkova}, \& {Kroupa}}]{Yan2017b}
{Yan}, Z., {Jerabkova}, T., \& {Kroupa}, P. 2017

\bibitem[{{Zapartas} {et~al.}(2017){Zapartas}, {de Mink}, {Izzard}, {Yoon},
  {Badenes}, {G{\"o}tberg}, {de Koter}, {Neijssel}, {Renzo}, {Schootemeijer},
  \& {Shrotriya}}]{Zapartas2017}
{Zapartas}, E., {de Mink}, S.~E., {Izzard}, R.~G., {et~al.} 2017, \aap, 601,
  A29

\bibitem[{{Zonoozi} {et~al.}(2016){Zonoozi}, {Haghi}, \&
  {Kroupa}}]{Zonoozi2016}
{Zonoozi}, A.~H., {Haghi}, H., \& {Kroupa}, P. 2016, \apj, 826, 89

\end{thebibliography}

\begin{appendix}

\section{Disagreements on the $m_{\mathrm{str, max}}$--$M_{\mathrm{ecl}}$ relation}\label{Appendix_mMrelation}

Some studies have been arguing against the existence of a $m_{\mathrm{str, max}}$--$M_{\mathrm{ecl}}$ relation and favor the stochastic star formation scenario. Partially such claims are based on unresolved observations with large uncertainty on cluster age and mass determinations. More importantly, a typical issue many papers have, for instance in \cite{Fumagalli2011}; \cite{Andrews2013} and \cite{Dib2017}, is to consider the $m_{\mathrm{str, max}}$--$M_{\mathrm{ecl}}$ relation as a simple truncation mass in a randomly sampled IMF (or ECMF)\footnote{As stated in the SLUG paper, see \citet[their appendix A]{daSilva2012}, and \citet[their section 5.2]{Dib2017}.}. Although it has been stated clearly in \cite{Weidner2014} that this is not the case, neither \cite{Weidner2006} nor the original IGIMF paper \citep{Kroupa2003} highlighted the idea of optimal sampling which was developed later \citep{Kroupa2013}. And the implementation of optimal sampling is mathematically more involved than drawing random numbers. Since this issue is still highly debated, here we explain it again.

The significant $m_{\mathrm{str, max}}$--$M_{\mathrm{ecl}}$ relation refers to the small scatter of the observational data, suggesting that one should apply a more deterministic sampling method like optimal sampling, which results in a deterministic list of masses of the sampled stars. Given an embedded cluster mass, $M_{\mathrm{ecl}}$, the optimal sampling constraint (Eq.~\ref{eq:1intMstar}) leads to a most massive star mass, $m_{\mathrm{str, max}}$, that automatically fulfills the empirical $m_{\mathrm{str, max}}$--$M_{\mathrm{ecl}}$ relation. This is different from a combination of truncation mass limit and random sampling. For example in our Fig.~\ref{fig:MmaxMecl} below, the optimally-sampled result (the thick black line) reproduces the $m_{\mathrm{str, max}}$--$M_{\mathrm{ecl}}$ relation and describes the observational data rather than being an upper bound. The scatter around our theoretical prediction is accounted for by large observational uncertainties, systematic uncertainties and small randomness due to additional parameters not known to the theory. If the $m_{\mathrm{str, max}}$--$M_{\mathrm{ecl}}$ relation is used as a truncation mass limit in cooperation with random sampling, all the sampled stars would be situated below the thick black line in Fig.~\ref{fig:MmaxMecl}, thus deviating from the observational data. So the counter-papers arguing against the $m_{\mathrm{str, max}}$--$M_{\mathrm{ecl}}$ relation actually disagree with the truncation-plus-random-sampling scenario which is not what \cite{Weidner2006} and \cite{Kroupa2003} imply\footnote{The word "truncation" did not appear in \cite{Weidner2006} and \cite{Kroupa2003}. It only appears in \cite{Schulz2015} to describe the upper integration limit as mentioned above in Sec.~\ref{sec: Model-OSGIMF}.} as has been already explained by \cite{Weidner2014}.

\section{The IGIMF for different assumptions}\label{Appendix}

IGIMFs are shown in Fig.~\ref{fig:metal_alpha3_change} for different SFRs and metallicities according to Eq.~\ref{eq:alpha_3_metal}, where the metallicity of the gas phase of a galaxy is taking into consideration.  Lower metallicity cases result in a more top-heavy IMF, but this metallicity dependence vanishes at low SFRs as the change in $\alpha_3$ has no effect if there aren't any massive stars.\\

Six IGIMF models for different assumptions on $\alpha_3$ and $\beta$ are shown in Fig.~\ref{fig:alpha3_change_beta_Weidner} to \ref{fig:alpha3_fix_beta_fix}. In the plot captions, "fixed $\alpha_3$" means $\alpha_3=2.3$, "changeable $\alpha_3$" means $\alpha_3$ follows Eq.~\ref{eq:alpha_3}, "fixed $\beta$" means $\beta=2$, "changeable $\beta$" means $\beta$ follows Eq.~\ref{eq:beta-SFR}, "Weidner $\beta$" means Eq.~\ref{eq:beta-SFR} is only applied when SFR $>1$ M$_{\odot}$/yr.

The "changeable $\alpha_3$" and "changeable $\beta$" case plotted in Fig.~\ref{fig:IGIMF_SFR_} is our fiducial model for which the resulting $\alpha_3^{\mathrm{gal}}$--SFR relation best reproduces the observational constrains as shown in Fig.~\ref{fig:alpha3SFR}. Other models are shown below:

\begin{figure}[hbt]
    \center
    \includegraphics[width=9cm]{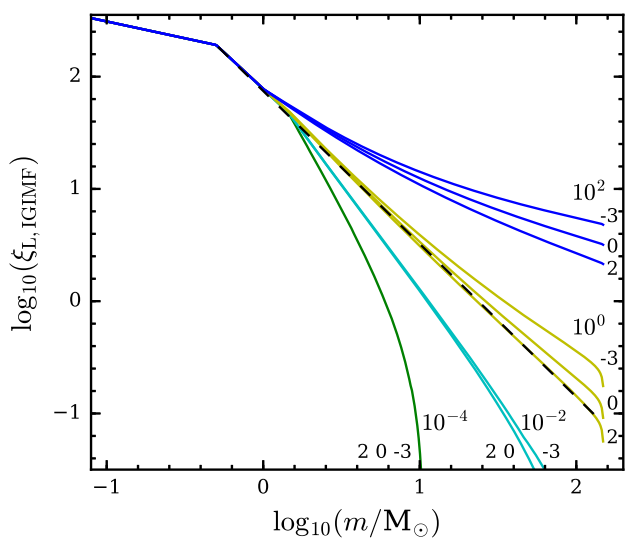}
    \caption{Same as Fig.~\ref{fig:IGIMF_SFR_} but using Eq.~\ref{eq:alpha_3_metal}, with \textcolor{green}{green}, \textcolor{cyan}{cyan}, \textcolor{yellow}{yellow}, and \textcolor{blue}{blue} lines for SFR = $10^{-4}$, $10^{-2}$, $10^{0}$, $10^{2}$ M$_{\odot}$/yr, respectively. There are three cases for each color with [Fe/H] = -3, 0, and 2 from upper to lower lines labeled in the plot, however, the three green lines overlap and the lower two cyan lines also overlap.}
    \label{fig:metal_alpha3_change}
\end{figure}

\begin{figure}[!hbt]
    \center
    \includegraphics[width=9cm]{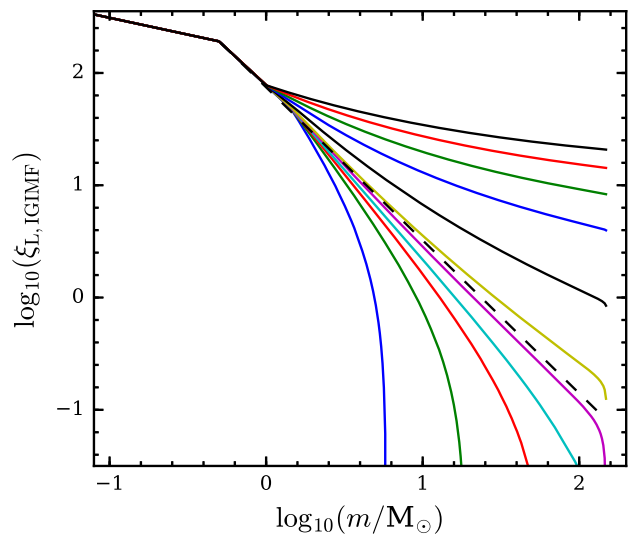}
    \caption{Same as Fig.~\ref{fig:IGIMF_SFR_} with changeable $\alpha_3$ and Weidner $\beta$.}
    \label{fig:alpha3_change_beta_Weidner}
\end{figure}
\begin{figure}[!hbt]
    \center
    \includegraphics[width=9cm]{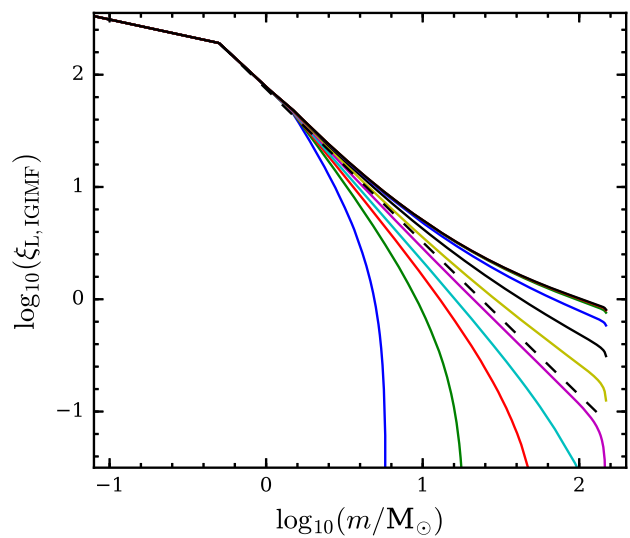}
    \caption{Same as Fig.~\ref{fig:IGIMF_SFR_} with changeable $\alpha_3$ and fixed $\beta$.}
    \label{fig:alpha3_change_beta_fix}
\end{figure}
\begin{figure}[!hbt]
    \center
    \includegraphics[width=9cm]{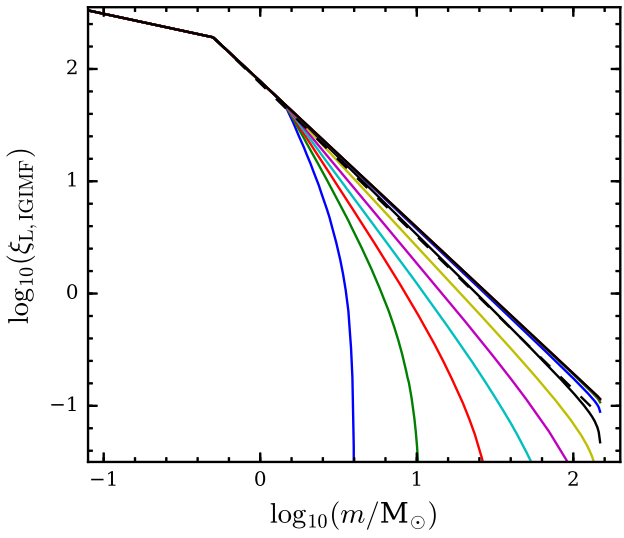}
    \caption{Same as Fig.~\ref{fig:IGIMF_SFR_} with fixed $\alpha_3$ and changeable $\beta$.}
    \label{fig:alpha3_fix_beta_change}
\end{figure}
\begin{figure}[!hbt]
    \center
    \includegraphics[width=9cm]{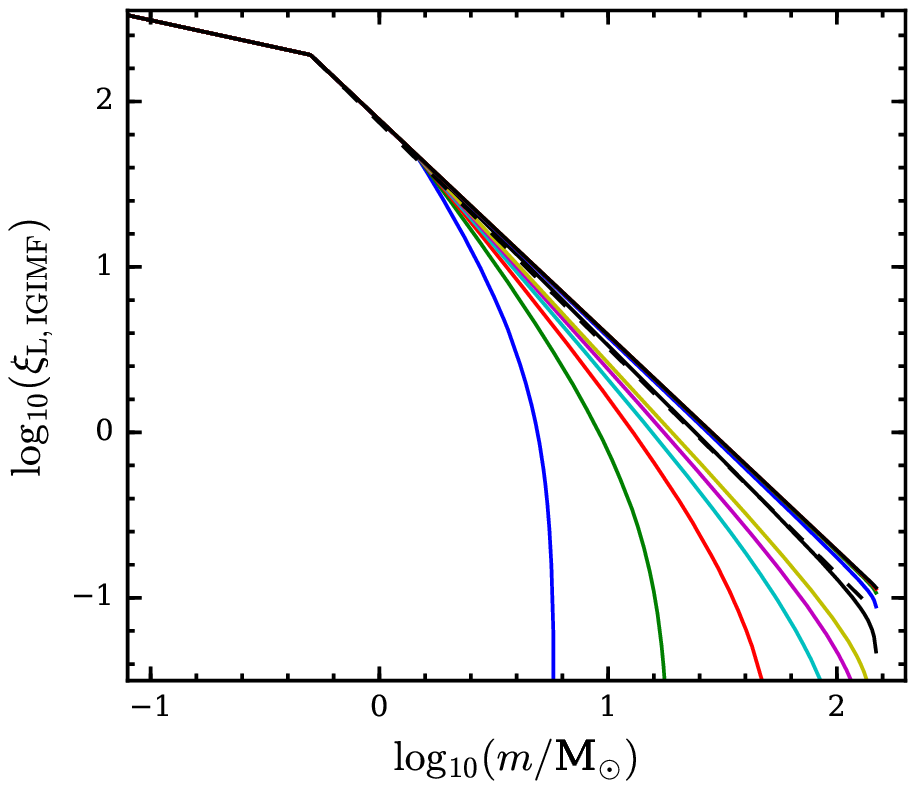}
    \caption{Same as Fig.~\ref{fig:IGIMF_SFR_} with fixed $\alpha_3$ and Weidner $\beta$.}
    \label{fig:alpha3_fix_beta_Weidner}
\end{figure}
\begin{figure}[!hbt]
    \center
    \includegraphics[width=9cm]{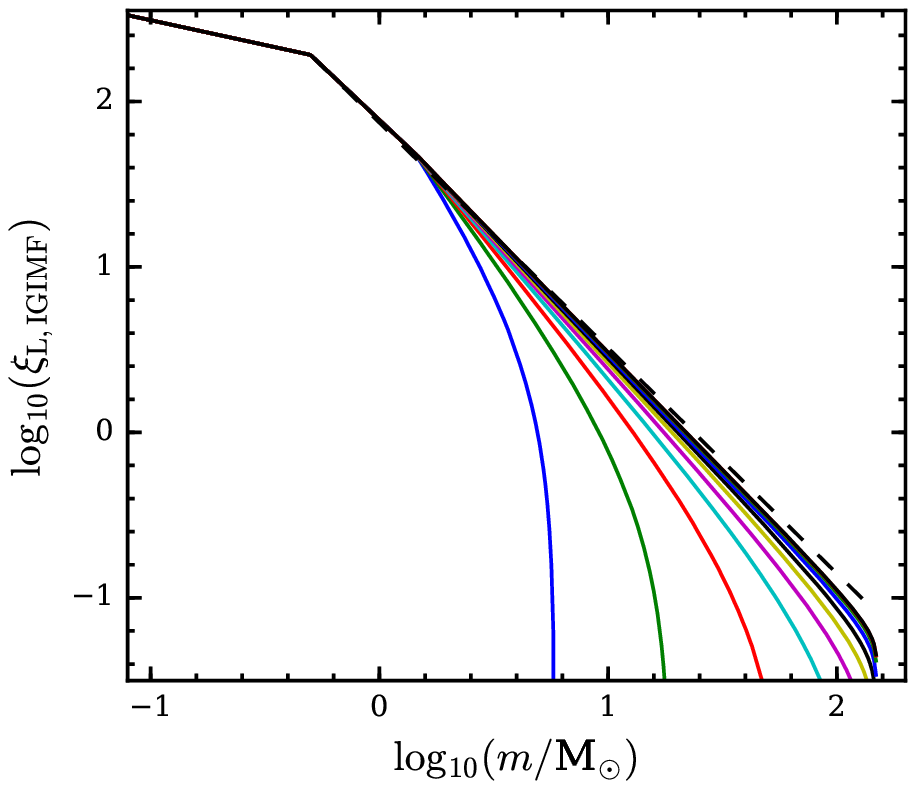}
    \caption{Same as Fig.~\ref{fig:IGIMF_SFR_} with fixed $\alpha_3$ and fixed $\beta$.}
    \label{fig:alpha3_fix_beta_fix}
\end{figure}

\section{The $m_{\mathrm{str, max}}$--$M_{\mathrm{ecl}}$ relation and the OSGIMF for a constant $\alpha_3$ assumption}\label{Appendix2}
\begin{figure}[!hbt]
    \center
    \includegraphics[width=9cm]{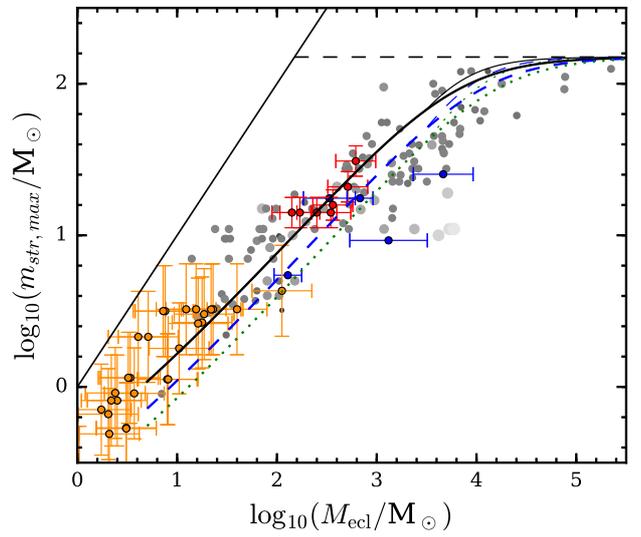}
    \caption{Same as Fig.~\ref{fig:MmaxMecl} but with the thick lines applying the fixed IMF assumption of $\alpha_3=2.3$ while the thin lines mark the line positions of Fig.~\ref{fig:MmaxMecl}.}
    \label{fig:MmaxMecl23}
\end{figure}

\begin{figure*}[!hbt]
    \center
    \includegraphics[width=\hsize]{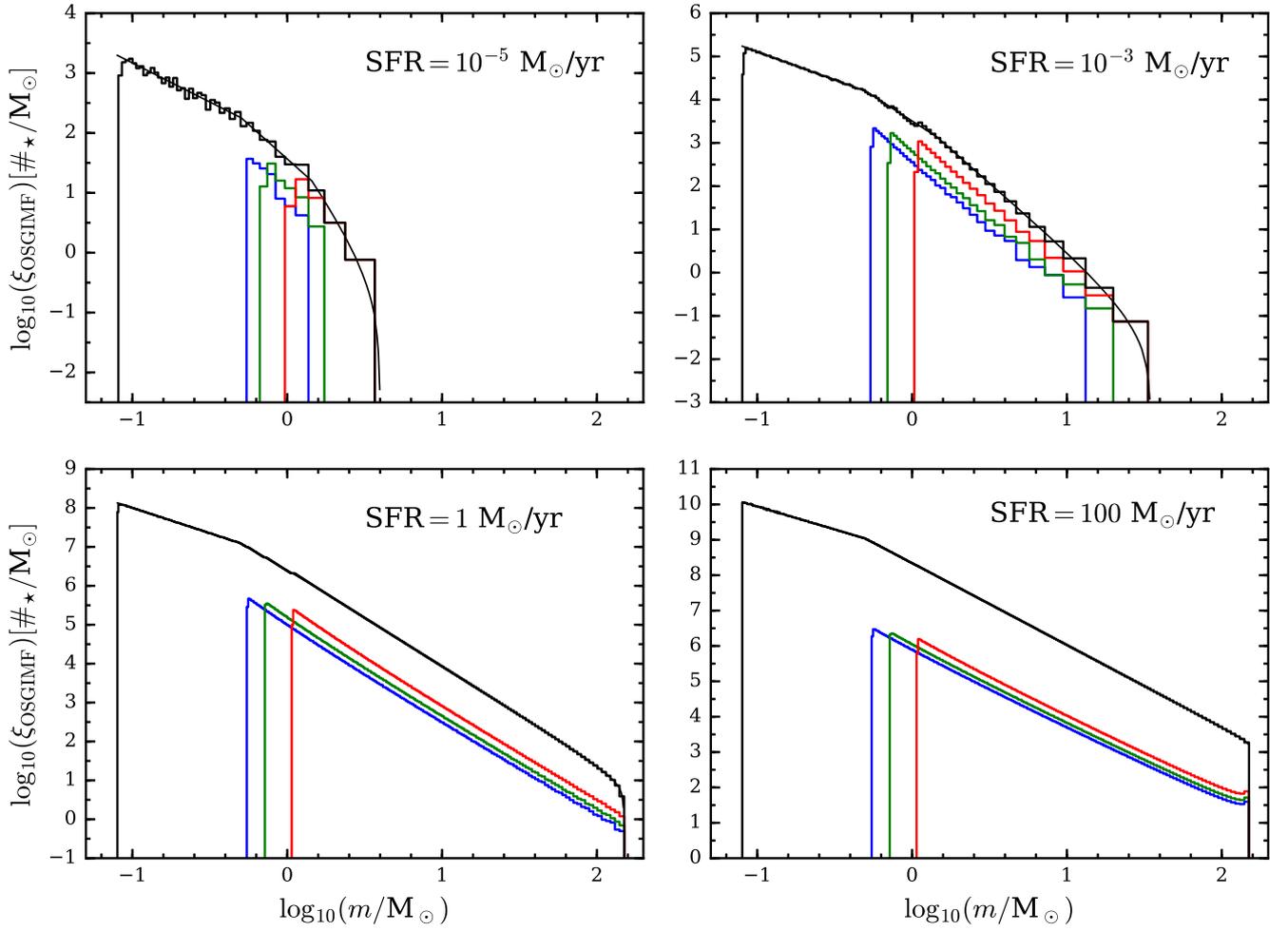}
    \caption{Same as Fig.~\ref{fig:OSGIMF} but with a fixed IMF assumption of $\alpha_3=2.3$. Notice the spoon feature in the lower panels of Fig.~\ref{fig:OSGIMF} disappears here.}
    \label{fig:OSGIMF2}
\end{figure*}

If we assume $\alpha_3=2.3$ is a constant instead of varying according to Eq.~\ref{eq:alpha_3_metal} and inherit everything else from Sec.~\ref{sec:Model-IGIMF}, the resulting $m_{\mathrm{str, max}}$--$M_{\mathrm{ecl}}$ relation would have no steepening feature in Fig.~\ref{fig:MmaxMecl}, as shown in Fig.~\ref{fig:MmaxMecl23}. This change also removes the spoon-feature in our Fig.~\ref{fig:OSGIMF}, as shown in Fig.~\ref{fig:OSGIMF2}.

\end{appendix}

\end{document}